\documentstyle[preprint,aps,floats,epsfig]{revtex}
\tighten

\def\half{\textstyle{1\over2}}
\def\quarter{\textstyle{1\over4}}
\newcommand{\be}{\begin{equation}}
\newcommand{\ee}{\end{equation}}
\newcommand{\bea}{\begin{eqnarray}}
\newcommand{\eea}{\end{eqnarray}}
\newcommand{\bml}{\begin{mathletters}}
\newcommand{\eml}{\end{mathletters}}

\begin{document}
\preprint{DTP/98/73, gr-qc/9810061}
\draft
%\tighten
\renewcommand{\topfraction}{0.8}
%%%%%%%%%%%%%%%%% HERE comment out next 2 lines
%@twocolumnfalse\endcsname
 
\title{Vortices and extreme black holes: the question of flux expulsion}
\author{Filipe Bonjour\footnote{E-mail: Filipe.Bonjour@durham.ac.uk},
	Roberto Emparan\footnote{E-mail: Roberto.Emparan@durham.ac.uk} and
        Ruth Gregory\footnote{E-mail: R.A.W.Gregory@durham.ac.uk}}
\address{Centre for Particle Theory, 
Durham University, South Road, Durham, DH1 3LE, U.K.
}

\date{\today}

\maketitle

\begin{abstract}
It has been claimed that extreme black holes exhibit a
phenomenon of flux expulsion for abelian Higgs vortices, irrespective
of the relative width of the vortex to the black hole. Recent work by
two of the authors showed a subtlety in the treatment of the event
horizon, which cast doubt on this claim. We analyse
in detail the vortex/extreme black hole system, showing that while flux
expulsion can occur, it does not do so in all cases. We give analytic proofs
for both expulsion and penetration of flux, in each case deriving a bound for
that behaviour. We also present extensive numerical work backing up, and
refining, these claims, and showing in detail how a vortex can end on a black
hole in all situations.  We also calculate the backreaction of the vortex on
the geometry, and comment on the more general vortex-black hole system.
\end{abstract}

\vspace{5mm}

\pacs{PACS numbers: 04.70.-s, 04.70.Dy, 11.27.+d \hfill DTP/98/73,
      gr-qc/9810061}

%%%%%%%%%%%%%%%%% HERE comment out next  line
%\vskip2pc]

%%%%%%%%%%%%%%%%%%%%%%%%%%%%%%%%%%%%%%%%%%%%%%%%%%%%%%%%%%%%%%%%%%%%%% 

\section{Introduction}

The story of black hole hair is an ongoing and interesting one. It was thought
for some time that black holes were relatively bland objects, classified by
very few parameters: charge, mass, and angular momentum. This picture has
changed significantly in the past decade with the discovery of various types of
solutions which carry other, more exotic, charges---such as the colored black
holes~\cite{BM}---and solutions with dressed horizons~\cite{LW}. What is clear
is that when a nontrivial topology is allowed for the matter fields, so-called
``no-hair'' theorems can often be evaded. In this paper, we are interested in
the question of abelian Higgs hair, which may occur because a U(1) vortex can
pierce, or even end, on a black hole~\cite{AGK}. This particular phenomenon is
interesting both from the point of view of hair for the black hole as well as
providing a decay channel for the disintegration of otherwise stable
topological defects~\cite{EHKT,HR,E1,GH,E2}.

Briefly, the results of~\cite{AGK} showed that is was possible for a vortex
solution of the U(1) abelian Higgs model---a Nielsen-Olesen
vortex~\cite{NO}---to thread a Schwarzschild black hole, and that the matter
fields reacted very little to the presence of the event horizon. Inclusion of
backreaction of the vortex on the geometry revealed that it was an appropriate
smooth version of the Aryal-Ford-Vilenkin geometry~\cite{AFV} discovered some
years previously. Further work~\cite{GH} showed that the conical singularities
in other more complicated geometries could be smoothed over by the vortex,
which allows the exact vacuum instanton for the splitting of a conical
defect~\cite{GP} to be used to construct a smooth instanton for splitting of
physical topological defects~\cite{EHKT,HR,E1}. A technical feature of these
smooth gravitational instantons was that they contained two U(1) fields, the
broken U(1) of the abelian Higgs vortex, and an unbroken U(1),
electromagnetism, needed to give a regular Euclidean section for the instanton
(although not needed for regularity of the Lorentzian section). Naturally this
raised the question as to whether the results of~\cite{AGK} were still valid in
the presence of this extra field. In the papers of Chamblin et.\
al.~\cite{CCES} it was argued that while the results of~\cite{AGK} were
qualitatively the same for nonextremal black holes, in the extremal limit a
completely new phenomenon occurred, and the flux of the vortex was expelled
from the black hole, rather like flux is expelled from a superconductor.

The evidence presented in~\cite{CCES} was of the form of analytic arguments for
high winding vortices, and numerical work representing a black hole/vortex
system for a variety of relative sizes of black hole to the string in which the
flux lines of the vortex appeared to consistently wrap the black hole.  In a
previous comment~\cite{BG}, two of us pointed out that there were some
difficulties with the numerical evidence as stated, and that while expulsion
was possible for thick, or high winding, vortices, it did not appear to occur
for thin vortices.  This would apparently solve a puzzle noted by Chamblin et.\
al., namely that one could, in principle, take a vortex terminating on a
near-extreme black hole (see~\cite{AG} for a detailed discussion of selection
rules for terminating vortices) and then charge the black hole up to
extremality. This would appear to be in contradiction with the principle of
flux expulsion.  However, if flux expulsion is not mandatory, then such a
puzzle never arises.  There is, however, another problem if flux is always
expelled. It appears that very thin vortices have a higher energy wrapping the
black hole than if there were no black hole present at all. This would mean
that the system is unstable and the string expels the black hole, which then
raises a physical paradox. A string outside a black hole reacts to the black
hole's gravitational field, therefore we might expect it to be attracted. On
the other hand, the vortex is not charged under the massless U(1), therefore it
has no reason to feel any repulsion, except for the putative flux expulsion
force. It is therefore not easy to see what the equilibrium solution of a
vortex-black hole system would be.  Of course this is a rather naive argument,
as it ignores any effects due to the conical deficit geometry of a gravitating
cosmic string, which does produce a repulsive force on charges \cite{Lin}, 
as well as an attractive force on masses \cite{S1}, 
a point which we return to in the conclusion after
we have explored the issue of gravitational backreaction of the vortex.

In this paper, we submit the abelian Higgs vortex/extreme black hole system to
an exhaustive analysis, with the intent of pinning down precisely when, or
indeed whether, flux expulsion can or cannot occur. We give analytic arguments
for flux expulsion in certain regions of parameter space, and flux penetration
in others, and back up the analysis with a wealth of numerical data. We discuss
the problems of numerical integration of this system (and how our work differs
from that presented in~\cite{CCES}) and how we have avoided these and ensured
accuracy of the integrations. We also consider in detail the vortex
terminating on the black hole. We then include a discussion of gravitational
backreaction before concluding.

\section{The Abelian Higgs Vortex}\label{sec:abh}

We start by reviewing the U(1) vortex in order to establish notation
and conventions. The action for an abelian Higgs system is
\be \label{abhact}
  S_1 = \int d^4x \sqrt{-g} \left [ D_{\mu}\Phi ^{\dagger}D^{\mu}\Phi -
        {\quarter} {\tilde G}_{\mu \nu}{\tilde G}^{\mu \nu} - {\quarter}\lambda
        (\Phi ^{\dagger} \Phi - \eta ^2)^2 \right ],
\ee
where $\Phi$ is a complex scalar field, $D_{\mu} = \nabla _{\mu} + ieB_{\mu}$
is the usual gauge covariant derivative, and ${\tilde G}_{\mu \nu}$ the field
strength associated with $B_{\mu}$.  We use units in which $\hbar=c=1$ and a
mostly minus signature. It is conventional to express the field content in a
slightly different manner in which the physical degrees of freedom are made
more manifest by defining real fields $X, \; \chi $ and $P_{\mu}$ by
\bml \bea
  \Phi (x^{\alpha}) &=& \eta  X (x^{\alpha}) e^{i\chi(x^{\alpha}) }  \\
  B_{\mu} (x^{\alpha}) &=& {1\over e} \bigl [ P_{\mu} (x^{\alpha}) - \nabla
    _{\mu} \chi (x^{\alpha}) \bigr ].
\eea \eml
These fields represent the physical degrees of freedom of the broken symmetric
phase; $X$ is the scalar Higgs field, $P_\mu$ the massive vector boson, and
$\chi$, being a gauge degree of freedom, is not a local observable, but can
have a globally nontrivial phase factor which indicates the presence of a
vortex. The existence of vortex solutions in the abelian Higgs model was argued
by Nielsen and Olesen~\cite{NO}, and in the presence of a vortex $\oint d\chi =
2\pi N$, where $N$ is the winding number of the vortex.

In terms of these new variables, the equations of motion are
\bml \label{vorteqn} \bea
  \nabla _{\mu}\nabla ^{\mu} X - P_{\mu}P^{\mu}X + {\lambda   \eta  ^2\over 2}
    X(X^2 -1) &=& 0 \\
    \nabla _{\mu}G^{\mu \nu} + {X^2 P^{\nu} \over \beta} &=& 0,
\eea \eml
where $\beta = \lambda/2e^2$ is the Bogomol'nyi parameter~\cite{B}, and
$G_{\mu\nu}$ is the field strength of $P_\mu$. 

\begin{figure}[htbp]
  \centerline{\epsfig{file=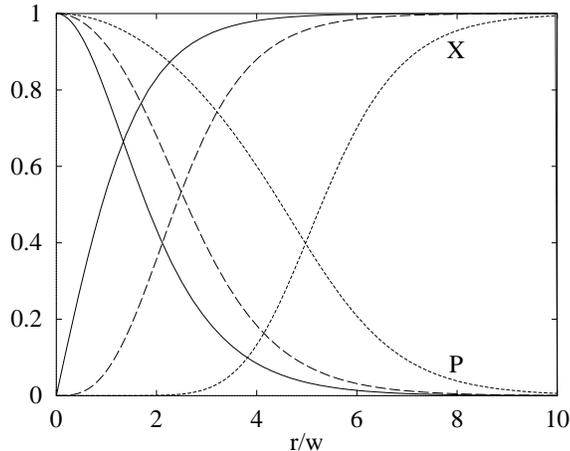,width=8.6cm}}
  \caption{The Nielsen--Olesen vortex for $\beta = 1$ and for $N = 1$ (solid
           lines), $N = 3$ (long-dashed lines) and $N = 10$.}
  \label{fig:NO}
\end{figure}

The simplest possible vortex solution is that in flat space:
\be
  X = X(R), \qquad P_\mu = NP(R)\partial_\mu\phi,
\label{xpform}
\ee
where $R=r\sqrt{\lambda}\eta$, $\{r,\phi\}$ are
polar coordinates, and
$X$ and $P$ satisfy the coupled second order ODE's
\bml \label{basic} \bea
  -X'' - {X'\over R} + {XN^2P^2\over R^2} + {\half} X(X^2-1) &=& 0 \\
  -P''+{P'\over R} + {X^2P\over\beta} &=& 0 
\eea \eml
For $N=1$, this is the Nielsen-Olesen solution, and gives an isolated vortex
for all $\beta$. The vortex core consists of two components---a scalar core
where the Higgs field differs from vacuum, roughly of width $1/\sqrt{\lambda}
\eta$, and a gauge core of thickness O($\beta^{1/2}/\sqrt{\lambda} \eta$). For
higher $N$, the solutions were given in~\cite{AGHK}, the principal differences
to $N=1$ being that the $X$-field is flattened ($X\sim R^N$) near the core, and
the string is correspondingly fattened. An additional difference is that for
$\beta>1$, higher winding strings are unstable to separation into $N$ unit
winding vortices~\cite{B}. Figure~\ref{fig:NO} presents some solutions for
$\beta = 1$.

In this paper however, we are interested in nontrivial solutions in curved
space, specifically in the presence of a charged black hole. This means that
our set-up now has three length scales, the two string core widths as already
mentioned, and the new scale---the black hole's outer horizon radius:
\bml \bea
  w_{\mbox{\scriptsize Higgs}} \sim m_{\mbox{\scriptsize Higgs}}^{-1} &=&
    \frac 1 {\sqrt{\lambda}\eta} \\
  w_{\mbox{\scriptsize gauge}} \sim m_{\mbox{\scriptsize gauge}}^{-1} &=&
    \frac 1 {\sqrt{2} e \eta} \\
  r_H &=& Gm + \sqrt{G^2m^2 - Gq^2}
\eea \eml
where $m$ is the ADM mass of the black hole, and $q$ its charge ($A_0 =
q/r$). We now set $\lambda \eta^2=1$, which considerably streamlines our
analysis and equations of motion. The gauge width of the core has already been
replaced by $\sqrt{\beta}$, and we now replace the mass and charge of the
Reissner-Nordstr{\o}m black hole by $M=Gm/w_H$ and $Q^2 = Gq^2/w_H^2$.  This
simply means that we have chosen to set the Higgs mass, rather than the Planck
mass, to unity. There are now no dimensionful quantities, and Newton's
constant, $G$, is a (small) number which we will represent by $\epsilon$, where
$\epsilon = 8\pi G \eta^2 $. For a GUT string $\epsilon = $O$(10^{-6})$, and
represents the gravitational strength of the string, which will be used in
section~\ref{sec:grav}, when we consider the gravitational backreaction of the
vortex.
 
For now, however, we ignore the gravitational backreaction of the vortex and,
as in \cite{CCES}, treat the vortex in the background Reissner-Nordstr{\o}m
geometry:
\bea \label{rnmetric}
  ds^2 &=& {\textstyle{\left ( 1 - {2M\over r} + {Q^2\over r^2} \right )}}
         dt^2 - {\textstyle{\left ( 1 - {2M\over r} + {Q^2\over r^2} \right
         )}}^{-1} dr^2 \nonumber \\ 
	&-& r^2 \left [ d\theta^2 + \sin^2\theta d\phi^2 \right ].
\eea
For the moment, it is irrelevant whether the geometry is the result of an
electric or magnetic potential since we only require the equations for the
vortex fields in the Reissner-Nordstr{\o}m {\it geometry}.
Substituting~(\ref{rnmetric}) in~(\ref{vorteqn}), and assuming the
form~(\ref{xpform}) for the $P_\mu$-field gives
\bml \label{genveq} \bea
  -{1\over r^2} \partial_r \left ( (r^2-2Mr+Q^2)\partial_r  X\right )
    -{1\over r^2\sin\theta} \partial_\theta \left ( \sin\theta \partial_\theta 
    X \right ) + {XN^2P^2\over r^2\sin^2\theta} + {\half} X(X^2-1) &=& 0 \\
  \partial_r \left [ \left ( 1 - {2M\over r} + {Q^2\over r^2} \right ) 
    \partial_r P \right ]  + {\sin\theta \over r^2 } \partial_\theta \left [
    {\partial_\theta P \over \sin \theta} \right ] - {X^2P\over \beta} &=& 0.
\eea \eml
In general, these equations are intractable analytically, however, as
in~\cite{AGK} we can extract some information in a particular limit---the `thin
string limit' where we assume $M\gg1$; the value of $Q$ is irrelevant. First
note that if we write $R = r\sin\theta$, and make the assumption that $X=X(R)$,
$P=P(R)$, then~(\ref{genveq}) becomes
\bml \label{rneq} \bea
  -X'' - {X'\over R} + {XN^2P^2\over R^2} + {\half} X(X^2-1)
    &=& {R^2\over r^2} \left ( {2M\over r} - {Q^2\over r^2} \right )
    \left [ X'' + {X'\over R} \right ] \\
  -P''+{P'\over R} + {X^2P\over\beta} &=& {R^2\over r^2} \left [
    {Q^2\over r^2} \left ( P'' - 2{P'\over R} \right ) - {2M\over r}
    \left ( P'' - {P'\over R} \right ) \right ].
\eea \eml
However, note that the RHS of each of these equations has the form $R^2/r^2$
times terms of order unity. Near the core $R$ is order unity, hence the RHS is
O($M^{-2}$), thus we have the flat space equations~(\ref{basic}) for which the
solutions are well known, and have the property that away from the core the
vortex fields tend to their vacuum values exponentially rapidly. Therefore, by
the time the premultiplying term in the RHS of~(\ref{rneq}) becomes
significant, the fields are essentially in vacuum, so any corrections will be
negligible. We can therefore regard the flat space solutions (as functions of
$R = r\sin\theta$) as a good approximation to the true solution in the thin
string limit. Note that this form of the solution pierces the horizon and does
not depend on the value of $Q$, therefore, using this argument, one would
expect that thin strings {\it always} penetrate the event horizon of a black
hole, whether or not it is extremal.

The argument developed so far starts from the thin string limit, but there is
another limit in which considerable information can be extracted analytically,
and which leads to the expectation that vortices are expelled from extremal
horizons. This regime, which can be regarded as a `thick string limit'
complementary to the one above, is attained for large winding number $N$. As
shown in~\cite{AGHK}, when $N$ is large the size of the vortex grows like
$\sqrt{N}$, and the unbroken phase inside the core is approached increasingly
faster, $X\sim R^N$. Consider then a black hole that sits well inside the
vortex core.  There, the field is expected to be very close to the symmetric
phase, so it seems reasonable to neglect the last term in~(\ref{genveq}b).
Then the equation can be solved by
\be
  P \approx 1-p(r^2 -Q^2) \sin^2\theta,
\ee
with $p\approx1/ (2N\sqrt{\beta})$~\cite{CCES}. From here we see that, in the
extremal limit in which the horizon is at $r=Q$, the magnetic flux across the
horizon, given by $G_{\theta\phi}=\partial_\theta P$, vanishes.  Moreover, it
is possible to solve for the Higgs field $X$ by setting $X=[b(r)\sin\theta]^N$,
and keeping only leading terms in $1/N$.  One finds, near the horizon,
\be
d(\log b) \propto {dr \over \sqrt{r^2 - 2M r +Q^2}}={dr \over 
\sqrt{(r-r_+)(r-r_-)}},
\ee
so that, if the black hole is extremal $(r_+=r_-)$ then $b \sim r-r_+$.  Hence,
$X$ vanishes as well on the extremal horizon.  Furthermore, a study of the
energy of the configuration shows that it is favourable for the black hole to
remain inside the vortex core~\cite{CCES}.

Indeed, the behavior of magnetic fields in the vicinity of extremal horizons
has been studied in more generality in~\cite{CEG} and the expulsion of the
flux---a phenomenon remarkably analogous to the Meissner effect in
superconductors---has been argued to be generic. Extremal horizons tend to
repel magnetic fields, at least if the latter are in, or approach, a phase of
unbroken symmetry, like in the core of the vortex. The argument above for
large $N$ vortices, however, is not fully conclusive, since it remains the
possibility that corrections of higher order in $M$ or in $1/N$ spoil the exact
expulsion phenomenon.

\section{Analytic arguments and bounds}

In the previous section we summarized the arguments of~\cite{CCES} in favour of
flux expulsion, and extended the thin string arguments of~\cite{AGK} which
seemed to indicate flux penetration. Both of these arguments appear compelling,
and we must examine the system closely to see what definite information can be
extracted. First, note that all methods are in agreement that there is flux
penetration for nonextremal black holes, therefore for the rest of this
section, we will only be considering extremal black holes, for which the metric
is
\be \label{xrnmetric}
  ds^2 = \left ( 1 - {M\over r} \right )^2 dt^2 -
    \left ( 1 - {M\over r} \right )^{-2} dr^2 - r^2 d\theta^2 -
    r^2\sin^2\theta d\phi^2.
\ee
This gives the vortex field equations
\bml \label{exteqs} \bea
  -{1\over r^2} \partial_r \left [ (r-M)^2 \partial_r  X\right ]
    -{1\over r^2\sin\theta} \partial_\theta \left ( \sin\theta \partial_\theta
    X \right ) + {XN^2P^2\over r^2\sin^2\theta} + {\half} X(X^2-1) &=& 0 \\
  \partial_r \left [ \left ( 1 - {M\over r} \right )^2
    \partial_r P \right ]  + {\sin\theta \over r^2 } \partial_\theta \left [
    {\partial_\theta P \over \sin \theta} \right ] - {X^2P\over \beta} &=& 0,
\eea \eml
which as before are not analytically soluble, however, we can extract
quite a bit of information about the solutions due to the nature of the
geometry near $r=M$.

The first observation is that if we set $r=M$, then the equations for the
horizon actually {\it decouple} from the exterior geometry:
\bml \label{horeqs} \bea
  -{1\over \sin\theta} \partial_\theta \left ( \sin\theta \partial_\theta
    X \right ) + {XN^2P^2\over \sin^2\theta} + {\half} M^2 X(X^2-1) &=& 0 
    \label{horxeq} \\
  \sin\theta\partial_\theta \left [{\partial_\theta P \over \sin \theta} \right
    ] - {M^2X^2P\over \beta} &=& 0, \label{horpeq}
\eea \eml
a phenomenon which does not occur in the nonextremal case. This means that the
vortex equations on the horizon are now ODE's and therefore easier to handle.
Note that the flux expulsion solution ($X=0$, $P=1$) always
solves~(\ref{horeqs}), therefore we cannot use any analysis of these equations
to demonstrate flux expulsion, but we can potentially show the nonexistence of
a penetration solution.

Therefore, assume that a piercing solution to the vortex equations does
exist throughout the spacetime, this means that a piercing solution must exist
on the horizon. 
This requires a nontrivial solution ($X(\theta), P(\theta)$) which is
symmetric around $\theta=\pi/2$ at which point $X$ has a maximum and $P$ a
minimum.  Let $X_m$ and $P_m$ be the extremal values of $X$ and $P$ attained.
In addition, expanding~(\ref{horeqs}) near the poles indicates that
$P_{,\theta}=0$ at the poles. Therefore, there exists a $\theta_0$ ($< \pi/2$)
such that $P_{,\theta\theta} = 0$ at $\theta_0$, and $P_{,\theta}(\theta_0) <
0$.

The structure of the proof is as follows; we use the properties of the 
solution at $\pi/2$ to derive an upper bound on $P$ and $P_{,\theta\theta}$
there. Then we use the behaviour of $P_{,\theta}$ to derive a lower bound
on $P_{,\theta\theta}$. At the very least, these must be consistent for a
piercing solution to exist, therefore, if the inequalities are
incompatible we conclude that the core solution is the only solution
on the horizon and therefore flux expulsion must occur.

To provide the upper bound, consider the $X$ equation at $\theta = \pi/2$.
Since $X_{,\theta\theta} < 0$, equation (\ref{horxeq}) implies $P_m^2 < {\half}
{M^2\over N^2} (1-X_m^2) < {\half} {M^2\over N^2}$ and hence
\be
  P_{,\theta\theta} (\pi/2) = {M^2\over\beta}
    X_m^2P_m < {M^3\over\sqrt{2} \beta N} X_m^2(1-X_m^2)^{1/2} < 
    {\sqrt{2} \over 3\sqrt{3}} {M^3\over\beta N}\;\;\; , \label{upper}
\ee
where the final inequality is obtained by maximizing over $X_m$.

For the lower bound, on the other hand note that at $\theta_0$ $|P_{,\theta}|$
takes its largest value, $|P_{,\theta_0}| = {M^2\over\beta} X^2P \tan\theta_0 <
{M^2\over\beta} \tan \theta_0$, and hence
\be
  {M\over\sqrt{2}N} > P_m > 1 - {\pi \over 2} |P_{,\theta_0}| \hspace*{1cm}
    \Rightarrow \hskip 1cm |P_{,\theta_0}| > {2\over\pi} \left ( 1 - {M\over
    \sqrt{2}N} \right ).
\ee
Assuming $M<\sqrt{2}N$, this gives ${\pi\over 2} - \theta_0 \leq \cot\theta_0 <
{\pi M^2 \over 2\beta \left ( 1 - {M\over\sqrt{2}N}\right )}$.  But then for
$M^2X_m^2<2\beta$ one can show that $P_{,\theta\theta}$ has a maximum at
$\pi/2$, hence
\be \label{lower}
  P_{,\theta\theta}(\pi/2) \geq {P_{,\theta}(\pi/2) - P_{,\theta} (\theta_0)
    \over \pi/2 - \theta_0} > \left ( {2\over\pi} \right )^2 {\beta\over M^2}
    \left ( 1-{M\over\sqrt{2}N} \right )^2.
\ee
Therefore, by comparing~(\ref{upper}) and~(\ref{lower}), we see that an
absolute minimum requirement for a piercing solution is the consistency of
these two bounds, i.e.
$$
  \sqrt{2} \pi^2 M^5 > 12\sqrt{3} \beta^2 N \left( 1-{M\over\sqrt{2}N} \right
    )^2.
$$

Turning this around therefore, and writing 
${\cal M} = {M\over\sqrt{2}N}$, we may conclude that
the vortex flux lines {\it must} be expelled from an extreme
Reissner-Nordstr{\o}m black hole if
\be \label{expbnd}
  {{\cal M}^5 \over (1-{\cal M})^2} < {3\sqrt{3} \over 2\pi^2} {\beta^2\over
    N^4} \simeq {\beta^2\over 4N^4}.
\ee
For $N=\beta=1$, this gives $M<0.7$; note that this is a rather weak bound, in
fact we would expect flux expulsion to be mandatory for $M$ somewhat in excess
of 0.7, but this method at least provides an analytic proof giving a definite
bound for $M$. Numerical work (section~\ref{sec:num}) actually places this
bound at about 1.9 (see figure~\ref{fig:horizon1}).

It is interesting to note the variation of this bound with $N$ and $\beta$.
For large $N$, ${\cal M}^5 < O(N^{-4})$, or $M < O(N^{1/5})$. This means that
the larger $N$, the larger the black hole can be and still have flux
expulsion. This is in agreement with the observation that large $N$ vortices
are thicker than their single winding number counterparts, therefore we would
expect flux expulsion to occur more readily. Indeed, this is how~\cite{CCES}
originally argued for flux expulsion.

For $\beta\to0$ we see that $M\leq O(\beta^{2/5})$, i.e.\ that flux expulsion
only occurs for extremely small black holes. To understand this, recall that
the fall-off of the $P$ field is $P\simeq e^{-R/\sqrt{\beta}}$ therefore as
$\beta\to0$, the magnetic flux core of the string is getting smaller as
$\sqrt\beta$. This is consistent with the above bound. It is therefore
interesting to look at large $\beta$, since in this limit, the magnetic core
becomes very diffuse, and we are left with the Higgs core, which is no longer
exponentially cut-off, but follows a power law cut-off approaching that of the
global string.  For large $\beta$, the bound~(\ref{expbnd}) gives $ {\cal M} <
1 - O(\beta^{-2})$, hence $M < \sqrt{2}N + O(\beta^{-2})$.  Therefore, for
small charge Higgs, or global strings, we expect flux expulsion to occur for
$M$ of order the winding number of the string.

Having shown that flux expulsion must occur for sufficiently thick strings,
what of the argument of the previous section, which appeared to indicate that a
thin string would pierce the event horizon? Obviously, since the core is always
a solution to the horizon system of equations, we cannot use an argument based
on this system to argue flux penetration, but instead we must look at the full
PDE system of equations in the exterior region of the horizon, equations
(\ref{exteqs}).

Now, assume that there is a flux expulsion solution, then on $r=M$, $X=0$ and
$P=1$. Therefore, near $r=M$, $M^2X^2 \ll 1$
and $[(r-M)^2X_{,r}]_{,r} >0$; hence
\be
  {\half} M^2 X\sin^2\theta + \sin\theta \partial_\theta \sin\theta
  \partial_\theta X < XN^2P^2 < XN^2 \label{nbdxeq}
\ee
in this region. Now, we know that $X$ is symmetric around $\pi/2$, peaking at
some maximum $X_m$, and also that $\sin\theta \partial_\theta X$ vanishes at
$0,\pi/2$ and $\pi$; therefore let $\theta_0$ be the value at which
$\partial_\theta \sin\theta \partial_\theta X=0$, which must satisfy $ {\half}
M^2 \sin^2\theta_0 < N^2 $.  If $M<\sqrt{2}N$, then this inequality is clearly
satisfied, so we now take $M>\sqrt{2}N$, and let $\alpha>\theta_0$ be defined
by $M^2\sin^2\alpha = 2N^2$.

Integrating~(\ref{nbdxeq}) on the range $(\theta,\pi/2)$, for $\theta>\alpha$
gives
\be
  X_{,\theta}(\theta) > X(\theta) \left [ {\half} M^2 \cot\theta
  + N^2 \csc\theta \ln \tan \theta/2 \right ]. \label{xlow}
\ee
But since $X_{,\theta\theta}<0$ on $[\theta_0,\pi/2]$, we can deduce $
X_{,\theta}(\theta) < {X(\theta) - X(\theta_0) \over \theta-\theta_0} <
{X(\theta)\over \theta - \alpha} $, hence for consistency with~(\ref{xlow})
\be
  {1\over N^2} > (\theta - \alpha) \left [ \csc^2\alpha \cot\theta
  + \csc\theta \ln \tan \theta/2 \right ]
\label{prcin}
\ee
must hold over the range $\theta \in (\alpha,\pi/2)$ for the expulsion
solution to hold. The actual bound on $M$ is then obtained by plotting these
curves and determining for which $M$ this inequality is always satisfied. For
$N=\beta=1$ we find that for $M^2 > 8.5$, this inequality is violated, hence
the vortex {\it must} pierce the horizon in this case.

For larger $N$, the lower bound on $M$ for a piercing solution to be forced
does increase, however, the ratio $M/N$ actually decreases. The LHS
of~(\ref{prcin}) is $N^{-2}$ and it is easy to see that this requires $\alpha =
\pi/2 - O(N^{-1/2})$. From this, we therefore obtain $M>\sqrt{2} N\left[1+
O(N^{-1}) \right]$.  Note that this argument does not depend on $\beta$. This
means that for large $\beta$ we still get piercing solutions for the same range
of $M$. Since $\beta\to\infty$ corresponds to the global string, this should
not be too surprising. For $\beta\to0$, although we expect some drop in the
value of $M$, this method is unable to detect this.

When the vortex radius is much larger than the black hole radius one can find
approximate explicit solutions for the fields near the horizon. Indeed, one can
construct these solutions for arbitrary winding number $N$, therefore
generalizing the solutions in~\cite{CCES}.

The solution for the magnetic field is readily found by noticing that if $M$ 
(or better, $M/\sqrt{\beta}$) is very small, then inside the vortex and close 
to the horizon the gauge field is well approximated by the solution to the 
massless field equation~\cite{CCES,CEG}
\be
  P \simeq 1- 2 M p (r-M) \sin^2\theta,
\ee
where $p$ is an integration constant equal to twice the magnetic field strength
at the center of the core. Next, by looking at the Higgs field equation, we can
see that close to the horizon the potential term ${1\over 2} X(X^2-1)$ is 
suppressed by a factor
$M^2$ which we are taking to be small. Therefore it can be neglected. After 
setting $P\approx 1$, the
resulting equation can be solved with $X$ a function exclusively of
$(r-M)\sin\theta$,
\be
  X \simeq k \left[ (r-M) \sin\theta \right]^N,
\ee
where $k$ is another integration constant. We see that on the horizon both the
gauge and the Higgs field are expelled. Comparison between these approximate
solutions and the results of the numerical calculations in the next section
shows good agreement.

\section{Numerical results}\label{sec:num}

To solve equations~(\ref{genveq}) numerically, we have used the technique
developed in~\cite{AGK}, which consists of relaxing initial configurations of
the fields $X$ and $P$ on the (rectangularly) discretized plane, $(r, \theta)
\to (r_i = r_+ + i dr, \theta_j = j d\theta)$. We therefore replace the fields
by their values on this grid, $X(r,\theta) \to X_{i,j} \equiv X(r_i, \theta_j)$
(and similarly for $P$), and the differential operators by suitably discretized
versions. Adopting the notation of~\cite{AGK} and~\cite{CCES} (that is, $X_{00}
= X_{i,j}, X_{\pm0} = X_{i\pm1,j}$ and $X_{0\pm} = X_{i,j\pm1}$), we find that
the discretized version of~(\ref{genveq}) is $X_{00} \to X_{00}^{\mbox{
\scriptsize new}}, P_{00} \to P_{00}^{\mbox{\scriptsize new}}$, where
\bml \label{genveqdis}\bea
  X_{00}^{\mbox{\scriptsize new}}
         &=& \frac{ \frac2r \left(1 - \frac Mr \right) \frac{X_{+0} -
            X_{-0}}{2\Delta r} + \frac{\cot\theta}{r^2} \frac{X_{0+} -
            X_{0-}}{2\Delta\theta} + \left(1 - \frac{2M}r +
            \frac{Q^2}{r^2} \right) \frac{X_{0+} + X_{0-}}{\Delta r^2} +
            \frac{X_{0+} + X_{0-}} {r^2 \Delta \theta^2}}%
	    {\left(1 - \frac{2M}r + \frac{Q^2}{r^2} \right) \frac2{\Delta
            r^2} + \frac2{r^2\Delta \theta^2} + \frac12\left(X_{00}^2 -
            1\right) + \left( \frac{N P_{00}}{r \sin\theta} \right)^2} \\
  P_{00}^{\mbox{\scriptsize new}}
          &=& \frac{\frac2{r^2} \left(M - \frac{Q^2}r \right) \frac{P_{+0} -
            P_{-0}}{2\Delta r} - \cot\theta \frac{P_{0+} - P_{0-}}{2r^2
            \Delta\theta^2} + \left(1 - \frac{2M}r + \frac{Q^2}{r^2}
            \right) \frac{P_{0+} + P_{0-}}{\Delta r^2} + \frac{P_{0+} +
            P_{0-}}{r^2 \Delta\theta^2}}%
            {\left(1 - \frac{2M}r + \frac{Q^2}{r^2} \right) \frac2{\Delta
            r^2} + \frac2{r^2 \Delta \theta^2} + \frac{X_{00}^2}{\beta}}.
\eea \eml

There is, however, a subtlety in this process: relaxation methods usually
require that the values of the fields be fixed at \emph{all} the boundaries of
the domain of integration, and although we know the asymptotic values of $X$
and $P$ at $r \to \infty$ (the vacuum) and at $\theta \to 0, \pi$ (the string
core values), the configuration at the horizon $r = r_+$ is in fact the main
result we expect from this numerical calculation. The solution to this problem
conceived in~\cite{AGK} was to update the values of the fields at the horizon
immediately after updating the interior of the grid. Note that this still
requires an initial guess for the fields on the horizon---a crucial point we
will return to later.  Replacing $r = r_+$ in~(\ref{genveq}), we obtain
equations on the horizon:
\bml \label{genveqhor} \bea
  \frac{r_+-r_-}{r_+^2} \left.\frac{\partial
    X}{\partial r}\right|_{r = r_+} &=& - \frac1{r_+^2 \sin\theta}
    \partial_\theta \left( \sin\theta \partial_\theta X \right) + \frac12 X
    \left(X^2 - 1\right) + \frac{N^2 X P^2}{r_+^2 \sin^2\theta} \\
  \frac{r_+-r_-}{r_+^2} \left. \frac{\partial
    P}{\partial r}\right|_{r = r_+} &=& - \frac{\sin\theta}{r_+^2}
    \partial_\theta \left( \frac{\partial_\theta P} {\sin\theta} \right) +
    \frac{X^2 P}{\beta};
\eea \eml
clearly, these equations reduce to~(\ref{horeqs}) in the extremal case.  We
discretize this in the same way that we discretized the equations on the
interior of the grid (except that we must now take discretized differential
operators that do not depend on $X_{-0}$ or $P_{-0}$). The resulting equations
are
\bml \label{genveqhordis} \bea
   X_{00} &\to& X_{00}^{\mbox{\scriptsize new}}
          = \frac{ \sqrt{M^2-Q^2} \frac{X_{+0}}{\Delta r} + \frac{X_{0+} +
	    X_{0-}}{2\Delta \theta^2} + \cot\theta \frac{X_{0+} -
            X_{0-}}{4\Delta \theta}}%
	    {\frac{\sqrt{M^2-Q^2}}{\Delta r} + \frac1{\Delta \theta^2} +
	    \frac{r_+}4 \left(X_{00}^2 - 1\right) + \frac12 \left(
	    \frac{NP_{00}} {\sin\theta} \right)^2} \\
   P_{00} &\to& P_{00}^{\mbox{\scriptsize new}}
          = \frac{\sqrt{M^2-Q^2} \frac{P_{+0}}{\Delta r} + \frac{P_{0+} +
	    P_{0-}}{2\Delta \theta^2} - \cot\theta \frac{P_{0+} -
            P_{0-}}{4\Delta \theta}}%
            {\frac{\sqrt{M^2-Q^2}}{\Delta r} + \frac1{\Delta \theta^2} +
   	    \frac{r_+}{2\beta} X_{00}^2}.
\eea \eml
The process of updating the interior of the grid and then the horizon at each
iteration was carried on until the modulus of the largest relative correction
on the grid became smaller than some $\varepsilon$:
\begin{equation}
  \max_{i,j} \left| \frac{X_{i,j}^{\mbox{\scriptsize new}} -
    X_{i,j}^{\mbox{\scriptsize old}}}{X_{i,j}^{\mbox{\scriptsize old}}}
    \right|,
    \max_{i,j} \left| \frac{P_{i,j}^{\mbox{\scriptsize new}} -
    P_{i,j}^{\mbox{\scriptsize old}}}{P_{i,j}^{\mbox{\scriptsize old}}} \right|
    < \varepsilon.
\end{equation}
($i$ and $j$ run over the entire grid, including the horizon.)

\begin{figure}[htbp]
  \centerline{\epsfig{file=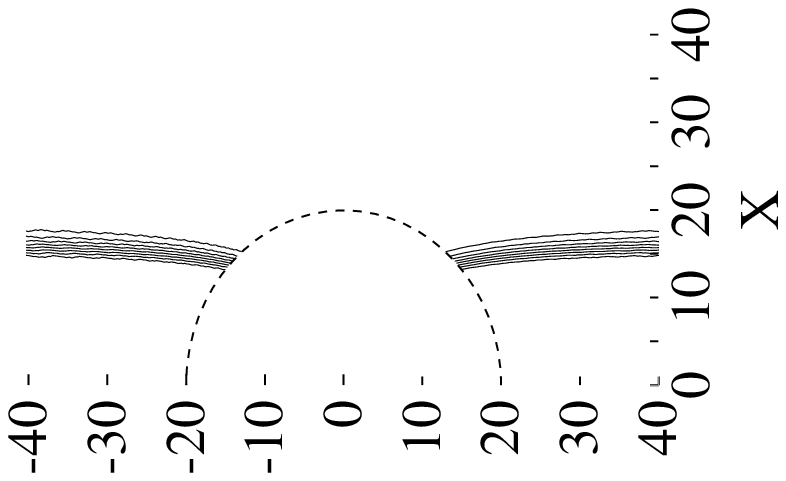,width=5cm,angle=270}~\qquad
              \epsfig{file=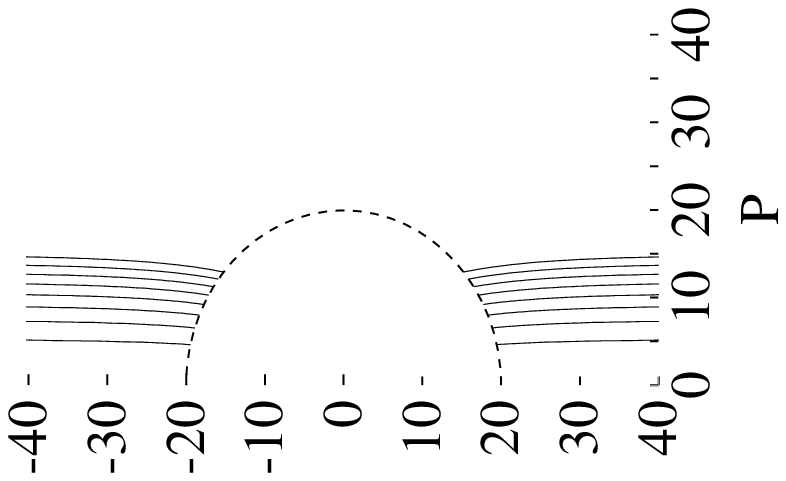,width=5cm,angle=270}}
  \caption{Contours of $X$ and $P$ for $M = 10, Q = 0$, $\beta = 1/2, N = 100$,
	   $\varepsilon = 10^{-4}$ and $N_r = N_\theta = r_m = 100$. (The
           dashed semi-ellipse represents the horizon.)}
  \label{fig:AGK3}
\end{figure}

The results obtained by our implementation of this method were compared with
the plots of~\cite{AGK}, and we found a satisfactory agreement; for instance,
figure~\ref{fig:AGK3} shows the contours of $X$ and $P$ for $M = 10, \beta =
1/2$ and $N = 100$; it can be directly compared with figure~3 from~\cite{AGK}.
($r_m$ is the maximum value for $r$, approximating $r \to \infty$.)

As in~\cite{AGK}, we have found that the thin string approximation
is excellent for thin vortices, and is a reasonable approximation 
even for thicker vortices, which tend to pinch slightly near the horizon.

We then turned our attention towards charged black holes, comparing now our
results with those of Chamblin et.\ al.~\cite{CCES}. We found, as they did,
that in nonextremal cases the picture remains qualitatively the same as for
uncharged black holes. For extremal black holes, however, our results differ
from their original claims. As reported in~\cite{BG}, we find that the claimed
expulsion of the matter fields for thin strings in this limit is the result of
a loophole in the numerical method (when applied to extremal black holes),
which does not take into consideration the decoupling of the horizon from the
main grid. [This was shown in Eq.~(\ref{horeqs}) and can also be seen
from~(\ref{genveqhordis}): if $Q = M$, these equations do not contain the terms
$X_{+0}$ or $P_{+0}$.] In this case, the core configuration is \emph{always an
exact solution} on the horizon; bearing in mind that the relaxation method
updates the fields from a user-supplied initial guess, we see that if the
initial guess made on the horizon is core, then the horizon will \emph{never}
be updated! In fact, \cite{CCES} always started from this guess on the
boundary, and therefore always obtained wrapping solutions in the extremal
case.

Obviously, this choice of initial conditions is important, moreover,
since we are dealing with a nonlinear system of PDE's, 
there is no reason for different initial configurations to relax to
the same solution. For this reason, we have considered the following three
initial data sets on the horizon:
\begin{itemize}
  \item \textbf{Core:} $X = 0, P = 1$,
  \item \textbf{Vacuum:} $X = 1, P = 0$,
  \item \textbf{Sinusoidal:} $X(\theta) = \sin\theta, P = 1$.
\end{itemize}
(The sinusoidal guess was chosen because it interpolates smoothly, and in a
simple and convenient way, the strings attached to the North and South poles
of the black hole.)

\begin{figure}[htbp]
  \centerline{\epsfig{file=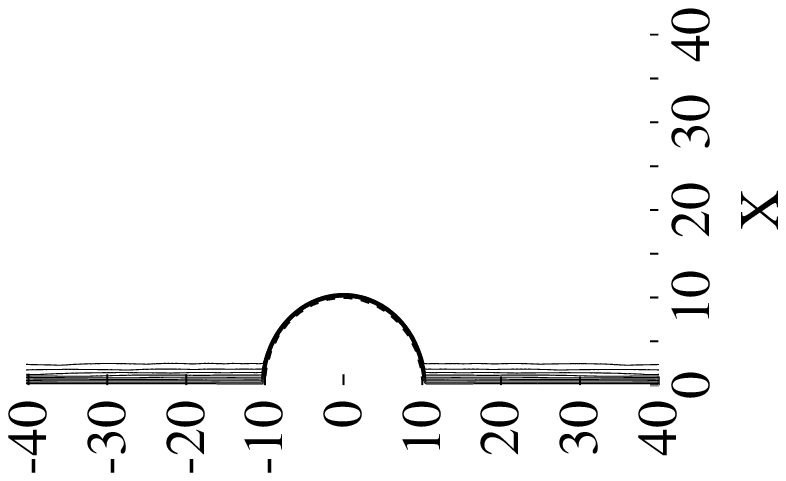,width=5cm,angle=270}~%
              \epsfig{file=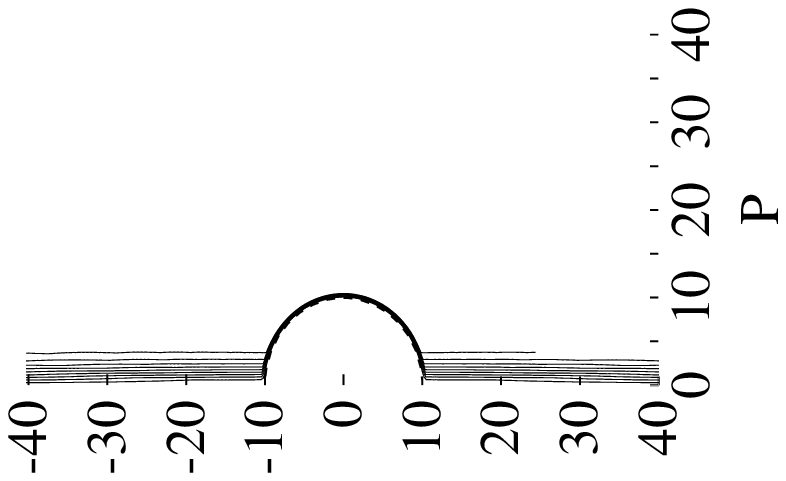,width=5cm,angle=270}}
  \centerline{\epsfig{file=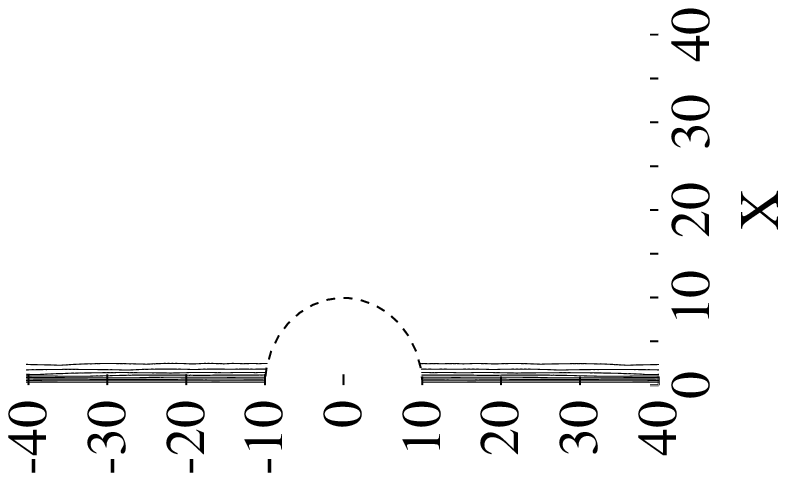,width=5cm,angle=270}~%
              \epsfig{file=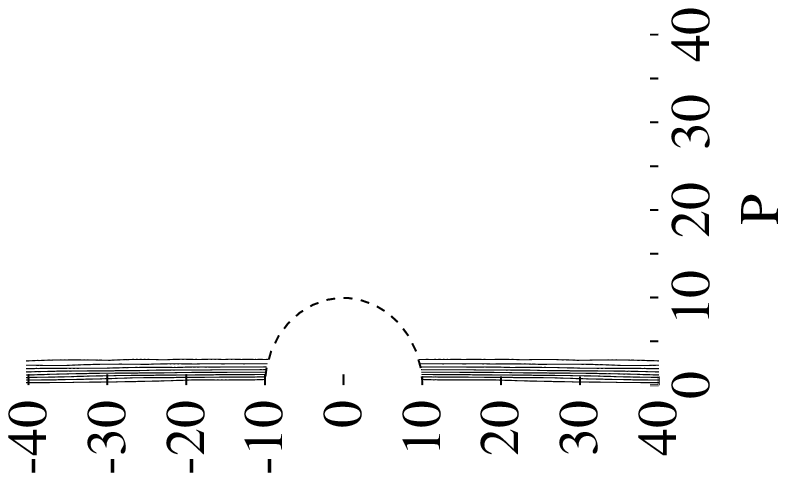,width=5cm,angle=270}}
  \caption{Contours of $X$ and $P$ for a core guess (top) and a vacuum (or
           sinusoidal) initial guess. The parameters are $M = Q = 10, \beta = N
           = 1, \varepsilon = 10^{-4}, N_r = N_\theta = 100$.}
  \label{fig:compic}
\end{figure}

Figure~\ref{fig:compic} displays the solutions obtained from the three initial
configurations. By computing and comparing the energy densities and total
energies of the fields on the grid, we were able to determine that \emph{inside
the grid} the three solutions were identical; on the horizon, however, the
solution relaxed from the core guess was a string core, whereas the solutions
obtained from the vacuum and the sinusoidal guesses were both the vacuum. The
comparison also showed that the wrapping solution had a higher total energy
than the piercing one; the difference is of course due to the jump of the
fields from the horizon to the interior of the grid. (Making the stepsize $dr$
smaller failed to smooth out this sharp jump.)

To summarize, we have three (physical and numerical) reasons to prefer the
piercing solution to the wrapping one for thin strings in the extremal limit:
it is smooth, numerically more robust, and energetically favourable.

To determine how the transition from a piercing to a wrapping solution occurs
as we thicken the string, we take advantage of the fact that, on the horizon,
we now have ODE's.  This allows for much quicker and more accurate numerical
methods; for the following calculations we have used the relaxation routine
\texttt{solvde} of Ref.~\cite[chapter 17]{NR}, starting from a vacuum guess.

\begin{figure}[htbp]
  \centerline{\epsfig{file=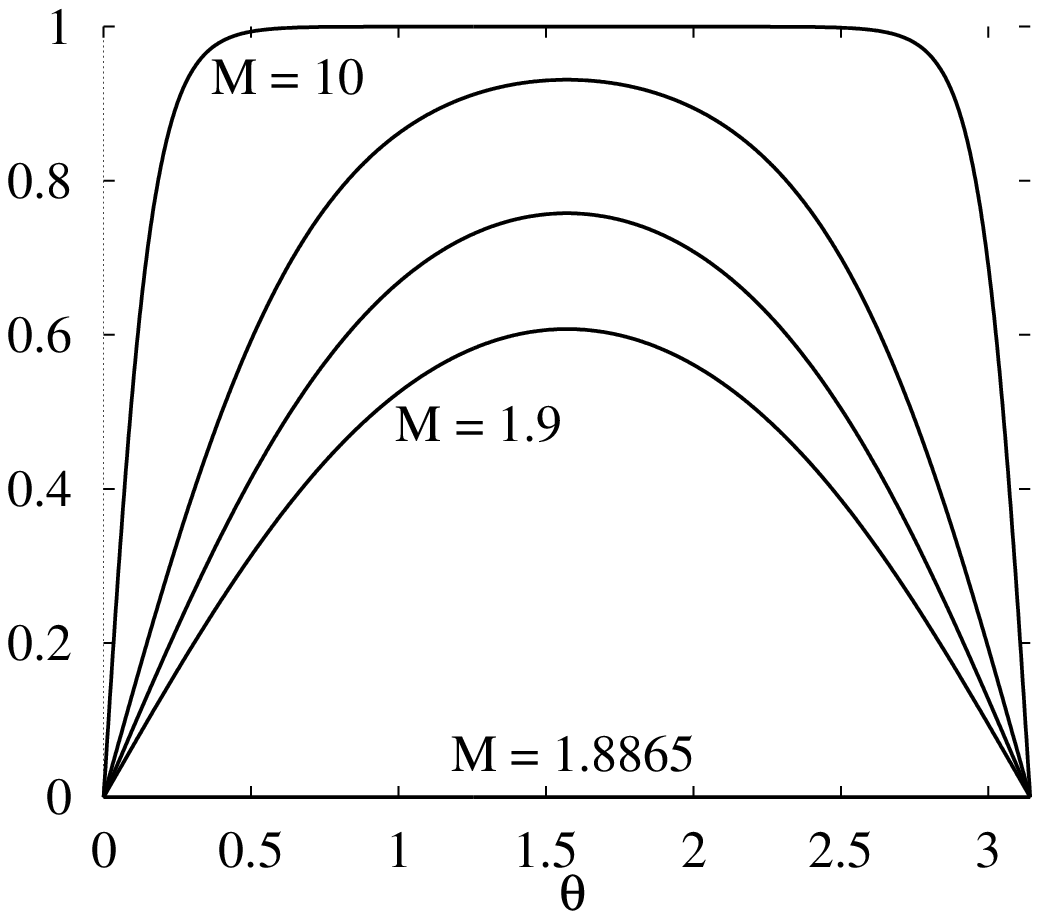,width=6.8cm} \qquad
  		\epsfig{file=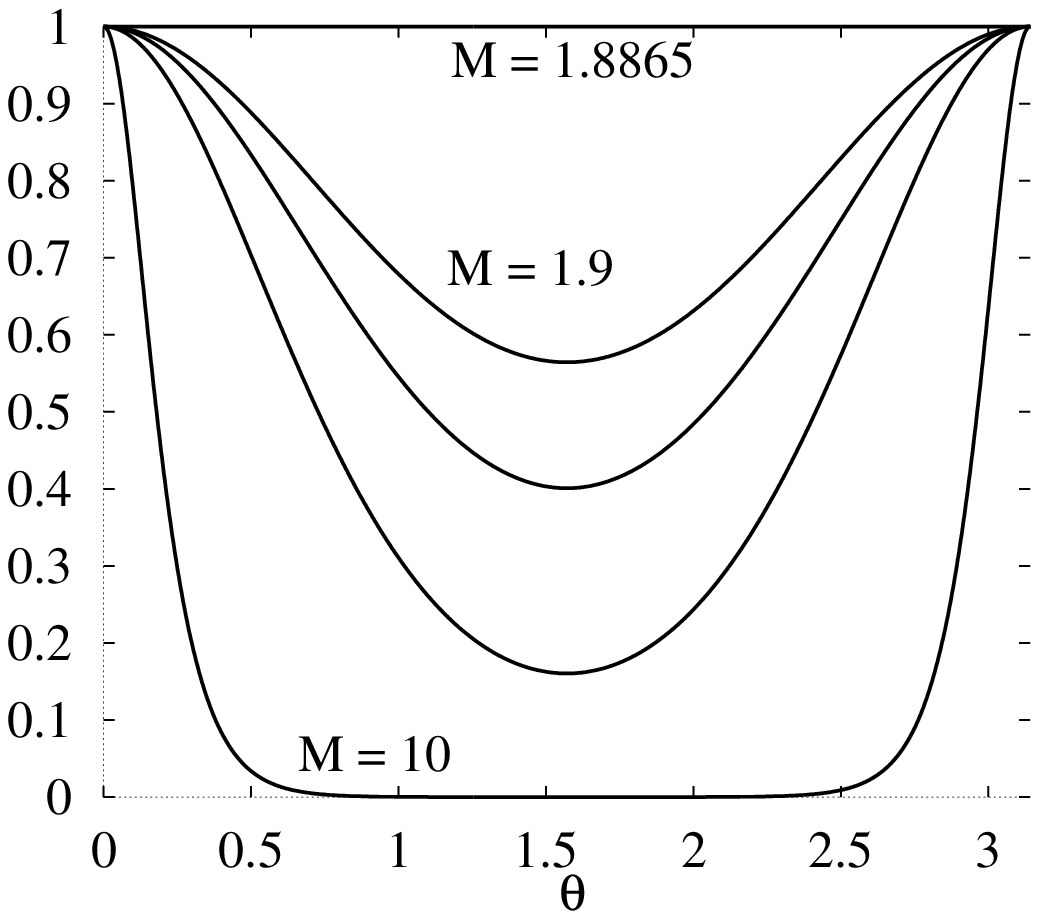,width=6.8cm}}
  \caption{Functions $X(\theta)$ (left) and $P(\theta)$ on the horizon
           for $\beta = N = 1$, a vacuum guess and $M = 10, 2.5, 2, 1.9$
           and $1.8865$.}
  \label{fig:horizon1}
\end{figure}

\begin{figure}[htbp]
  \centerline{\epsfig{file=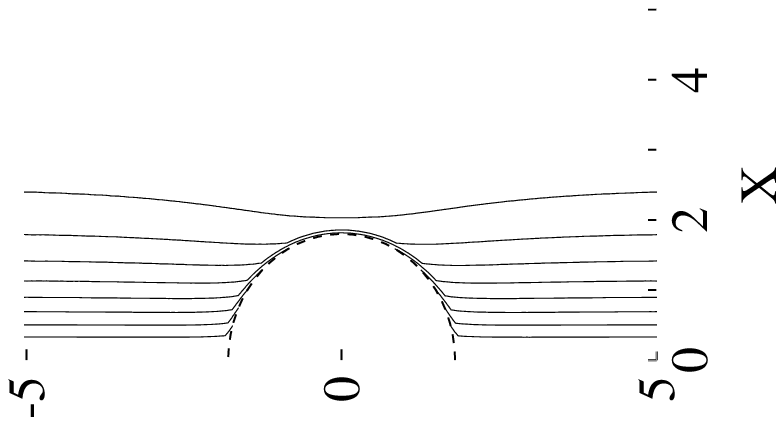,width=5cm,angle=270} \qquad
              \epsfig{file=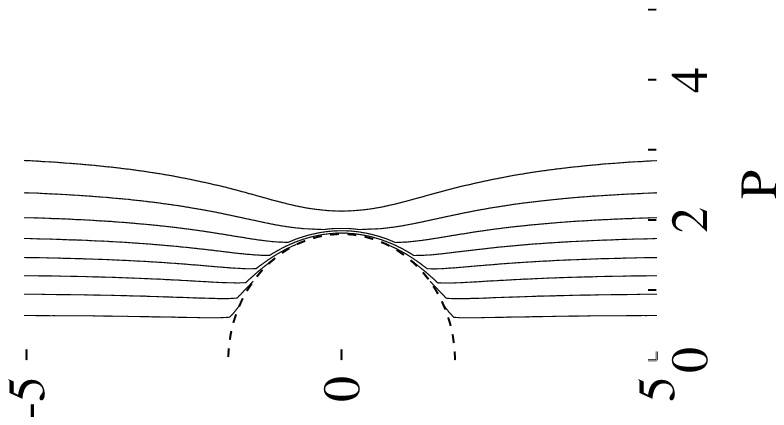,width=5cm,angle=270}}
  \caption{Contours of $X$ and $P$ for $M = Q = 1.8,$ $N = \beta = 1,
           \varepsilon = 10^{-3}$ and $N_r = N_\theta = 100$.}
  \label{fig:expell}
\end{figure}

\begin{figure}[htbp]
  \centerline{\epsfig{file=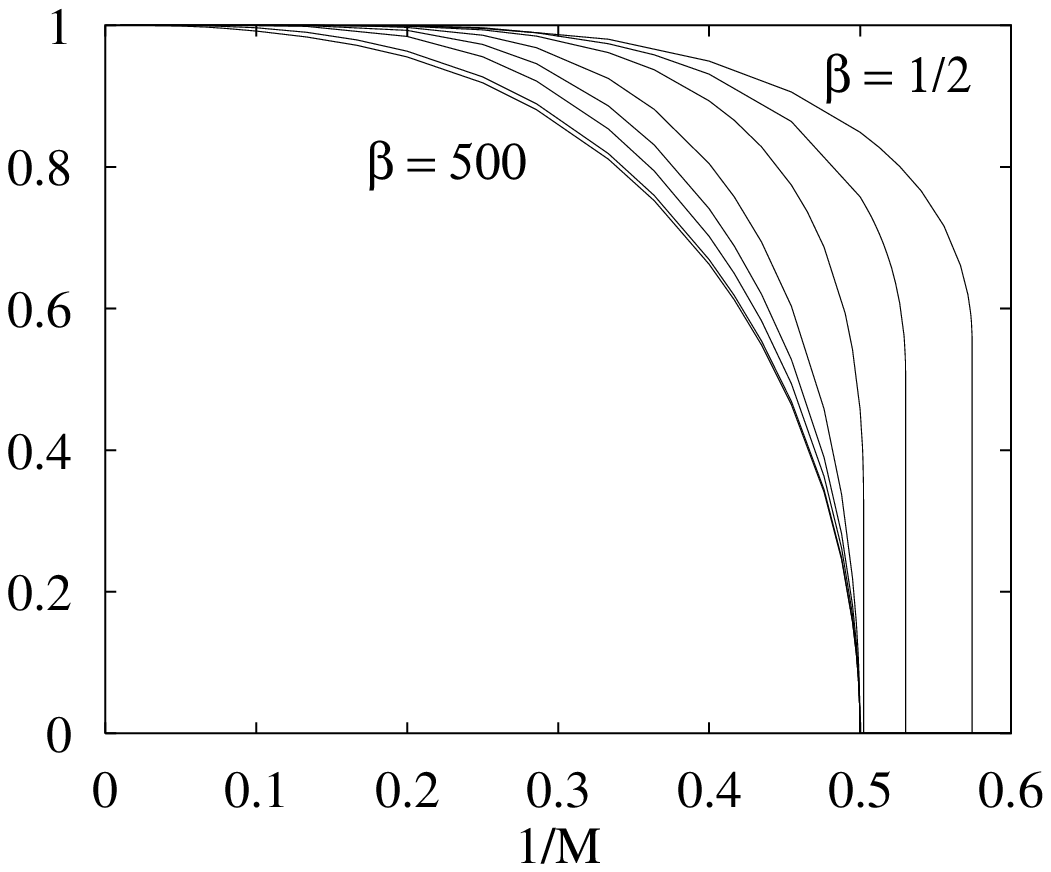,width=6.8cm} \qquad
              \epsfig{file=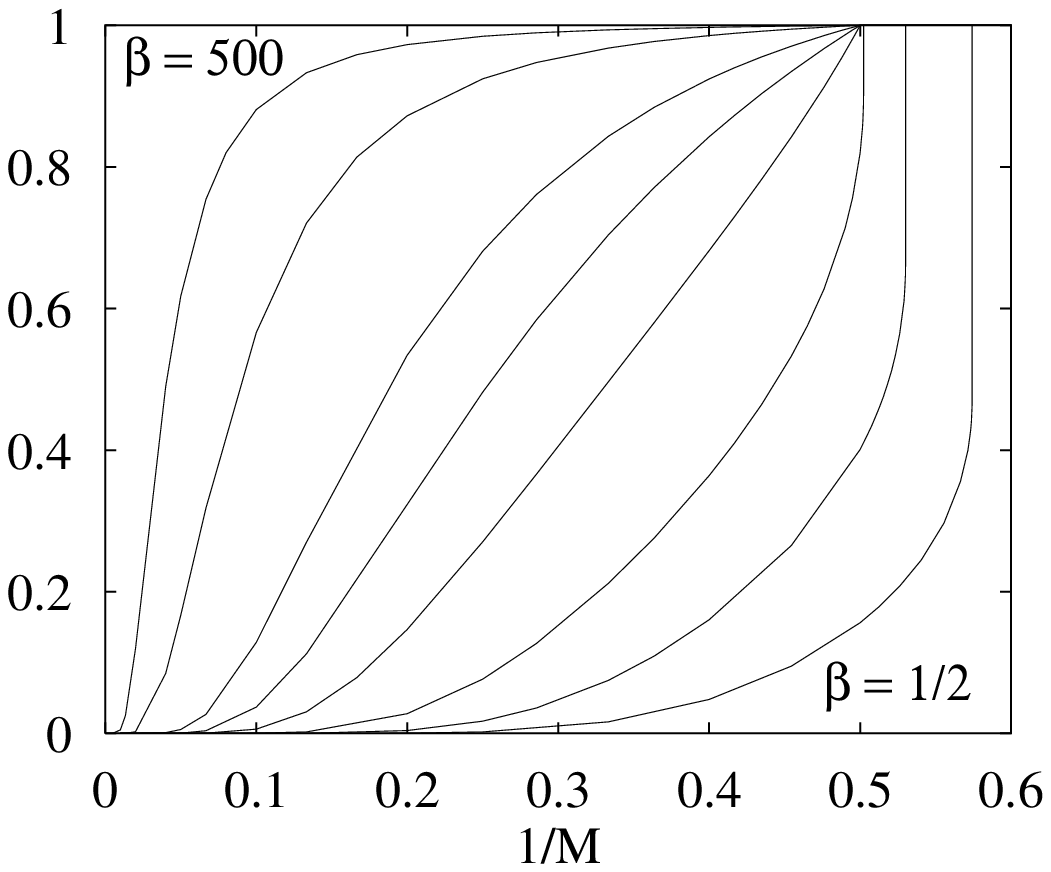,width=6.8cm}}
  \caption{Values of $X(\pi/2)$ (left) and $P(\pi/2)$ as functions of the
           inverse energy of the extremal black hole. This is done for $N = 1$
           and the following values of $\beta$: 500, 100, 20, 10, 5, 2, 1 and
           1/2.}
  \label{fig:maxmin}
\end{figure}

Figure~\ref{fig:horizon1} shows the solutions on the horizon for $\beta = N =
1$. For massive black holes (or, equivalently, thin strings), the fields adopt
a vacuum profile on most of the horizon (symmetrically about the equator
$\theta = \pi/2$) and interpolate smoothly to their fixed core values on the
poles. The shapes of $X$ and $P$ remain the same for all values of $M$: $X$ has
non-vanishing $\theta$-derivatives at the poles, and a maximum at the equator;
$P$ has zero derivatives at the poles, and a minimum at the equator. As we
thicken the string, the maximum of $X$ and the minimum of $P$ move away from
the vacuum values. For $N = \beta = 1$ this transition is gentle at first, but
accelerates suddenly, as if the string had crossed a critical width at which it
is not able to pierce the horizon any more (see figure~\ref{fig:expell} for an
example of expelled solution). This can be observed on figure~\ref{fig:maxmin},
which shows the evolution of $X(\pi/2)$ and $P(\pi/2)$ with the inverse energy
of the black hole for a variety of values of $\beta$. Increasing $\beta$, as we
have remarked above, means thickening the $P$ tube, and we would therefore
expect a heavier black hole to be required to expel this flux. This is indeed
what we observe and, naturally, the curve for $P$ is more affected by this than
that for $X$.

\begin{figure}[htbp]
  \centerline{\epsfig{file=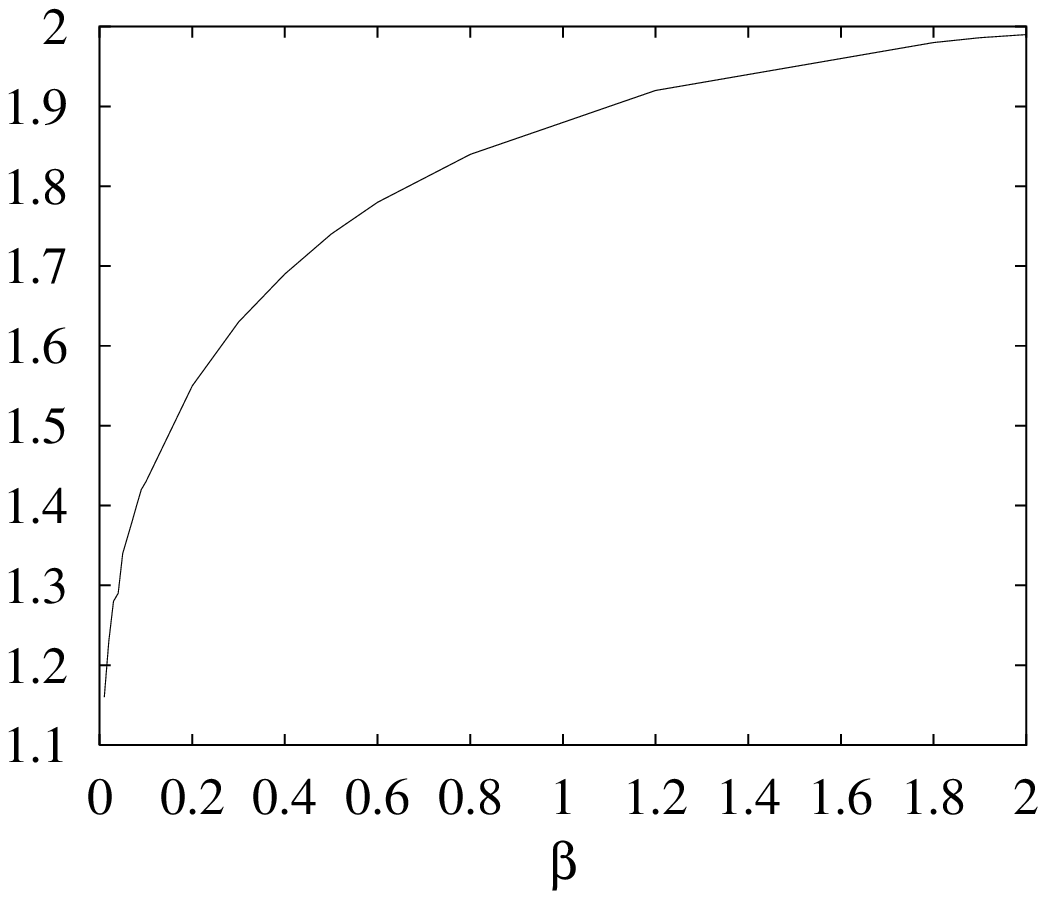,width=6.8cm} \qquad
              \epsfig{file=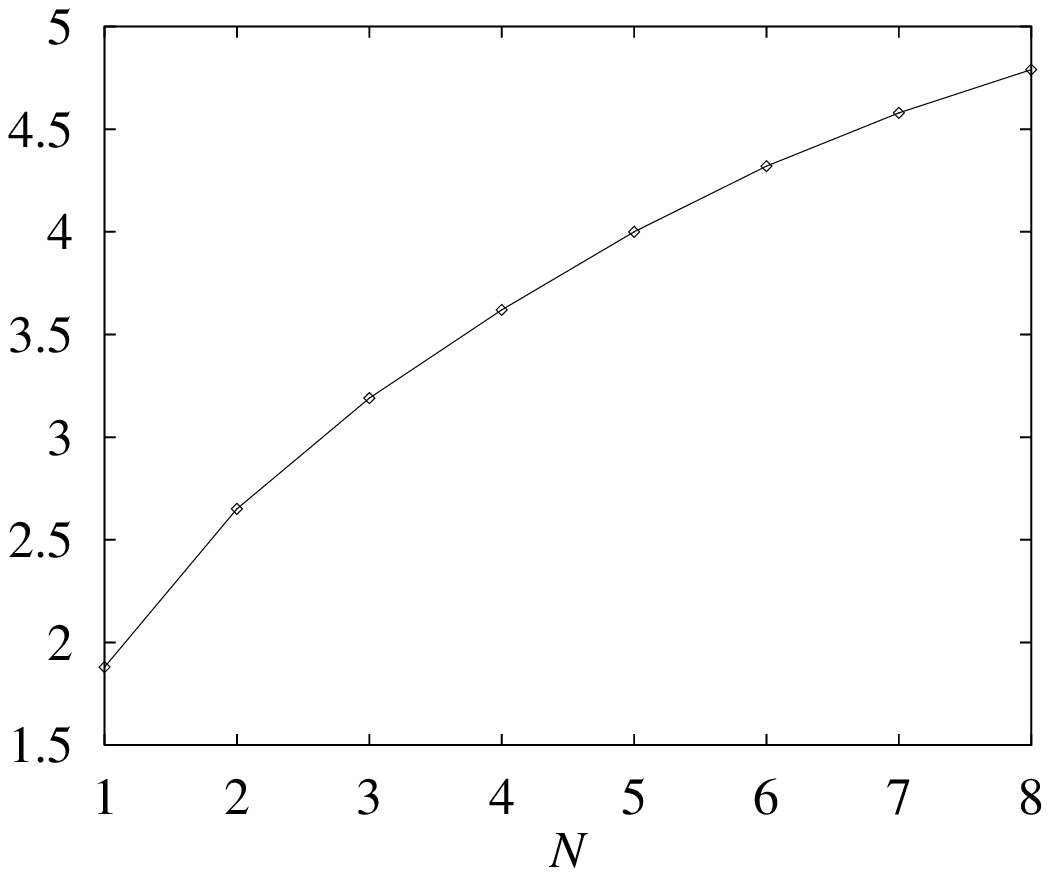,width=6.8cm}}
  \caption{Values of the `critical mass' $M_c$ at which the only solution on
           the horizon corresponds to the string being expelled, in function of
           $\beta$ (for $N=1$) and $N$ (for $\beta=1$).}
  \label{fig:Mcritic}
\end{figure}

The same shift of the curves towards higher values of $M$ happens when one
increases $N$, as anticipated from the fact that this also thickens the
string. Figure~\ref{fig:Mcritic} shows the evolution of the critical mass $M_c$
(defined to be that at which the horizon cannot support a penetrating solution
any more) in function of $\beta$ (for $N=1$) and of $N$ (for $\beta = 1$).

\section{String ending on a black hole}

Next, we turn our attention to the case of a string ending on the black
hole. This is an important configuration to consider, since it is the main
`phenomenological' input to the instantons mediating defect
decay~\cite{EHKT,HR,E1,GH,E2}. Originally,~\cite{CCES} supposed that such a
configuration may not be able to exist, however, the thin string arguments
indicate that at least for large $M$, such a configuration is possible. What we
will show is that while it is \emph{always} possible for a vortex to end on a
black hole, for small $M$ there is also a phenomenon analogous to flux
expulsion: the $X$ field is forced to sit in its unbroken phase ($X=0$) on the
horizon, and the $P$ field takes the form of a monopole potential.

To see this analytically, consider the horizon equations~(\ref{horeqs}). These
will have the boundary conditions $X=0$, $P=1$ at $\theta=0$, and $X=X_m$,
$P=0$ at $\theta = \pi$. The equation of motion for $P$,~(\ref{horpeq}), can
be integrated to show that
\be \label{pbd}
  \lambda(1+\cos\theta) < 2P < (1+\cos\theta)
\ee
where $\lambda = 1 - M^2X^2_m/\beta$. This shows that as $M^2/\beta\to0$, $P$
approaches its monopole form, $P_{_{\rm mon}}=\half (1+\cos\theta)$.  {}From
now on, we will assume that $M^2/\beta, M^2/N^2\ll1$ and explore the possible
solutions for $X$.  First of all, note that $X\equiv 0$, $P \equiv P_{_{\rm
mon}}$ is always a solution to the horizon equations, and it is this that we
will call the expulsion solution. Now suppose that a `piercing' solution
exists, then~(\ref{horxeq}) implies that $X$ has a local maximum at $\pi$,
which we will denote $X_m$. Defining $\theta_0$ by $\partial_\theta(\sin\theta
X_{,\theta})=0$, we see that at $\theta_0$, $P_0^2 = {M^2\over
2N^2}\sin^2\theta_0 (1-X_0^2)$.  Since $M/N$ is assumed small, $\theta_0$ will
obviously be close to $\pi$, and using the bounds on $P$, it is easy to see
that
\be
  {2\over \pi} (\pi-\theta_0) < \sin\theta_0<{4M\over \sqrt{2}N\lambda}
  \label{thobd}
\ee
and so $P$ will be extremely small. Integrating~(\ref{horxeq}) between $0$ and
$\pi$ then gives
\be
  \int_{\theta_0}^\pi X \left [ 
  {M^2\over 2N^2} \sin\theta(1-X^2) - {P^2\over\sin\theta}\right ]
  = \int_0^{\theta_0} X \left [
  {P^2\over\sin\theta} - {M^2\over 2N^2} \sin\theta(1-X^2)\right ]
  \label{sat}
\ee
The LHS of this equation can be readily be bounded above using~(\ref{thobd})
\be
  \int_{\theta_0}^\pi X \left [ 
  {M^2\over 2N^2} \sin\theta(1-X^2) - {P^2\over\sin\theta}\right ]
  < {4M^4X_m\over N^4\lambda^2}
\ee

The RHS is a little more tricky to bound below, but noting that
$X_{,\theta\theta}$ is positive on $[\pi/2,\theta_0]$, yet negative
($X_{,\theta\theta}= -M^2X_m(1-X_m^2)/4$) at $\pi$, we can bound $X_{,\theta}$
on $[\pi/2,\pi]$ by
\be
  X_{,\theta} < X_m N^2 \sin^3\theta_0
\ee
and hence we can bound $X$ below on $(\pi/2,\pi)$ by
\be
  X(\theta)
  > X_m \left [ 1 - {4\pi M^3\over N\lambda^3} \right ]
\ee
therefore
\bea
  \int_0^{\theta_0} X \left [
    {P^2\over\sin\theta} - {M^2\over 2N^2} \sin\theta(1-X^2)\right ]
    &>& \int_{\pi/2} ^{\theta_0} X\left 
    [ {P^2\over\sin\theta}- {M^2\sin\theta\over2N^2}\right]
    \nonumber \\
  &>& X_m {\lambda^2\over16} \left [ 1 - {4\pi M^3\over N\lambda^3} \right ]
    \left ( 1 - {8M^2\over N^2\lambda^2} \right )^2.
\eea
Comparing these bounds on the RHS and LHS of~(\ref{sat}) we see that an
expulsion solution is the only possible solution for
\be
  M^4 < {N^4 \lambda^4\over 64} \left ( 1 - {4\pi M^3\over N\lambda^3}\right )
  \left ( 1 - {8M^2\over N^2\lambda^2} \right )^2
\ee
For $N=\beta=1$ this gives $M<0.3$. As before, this is a very weak bound,
however, the important thing is that it shows that there is a lower bound on
the values of $M$ for which the $X$-field can vary on the horizon.  For $M$
sufficiently small, the Higgs field is forced to lie in its symmetric phase on
the horizon, and we get an expulsion solution.

\begin{figure}[htbp]
  \centerline{\epsfig{file=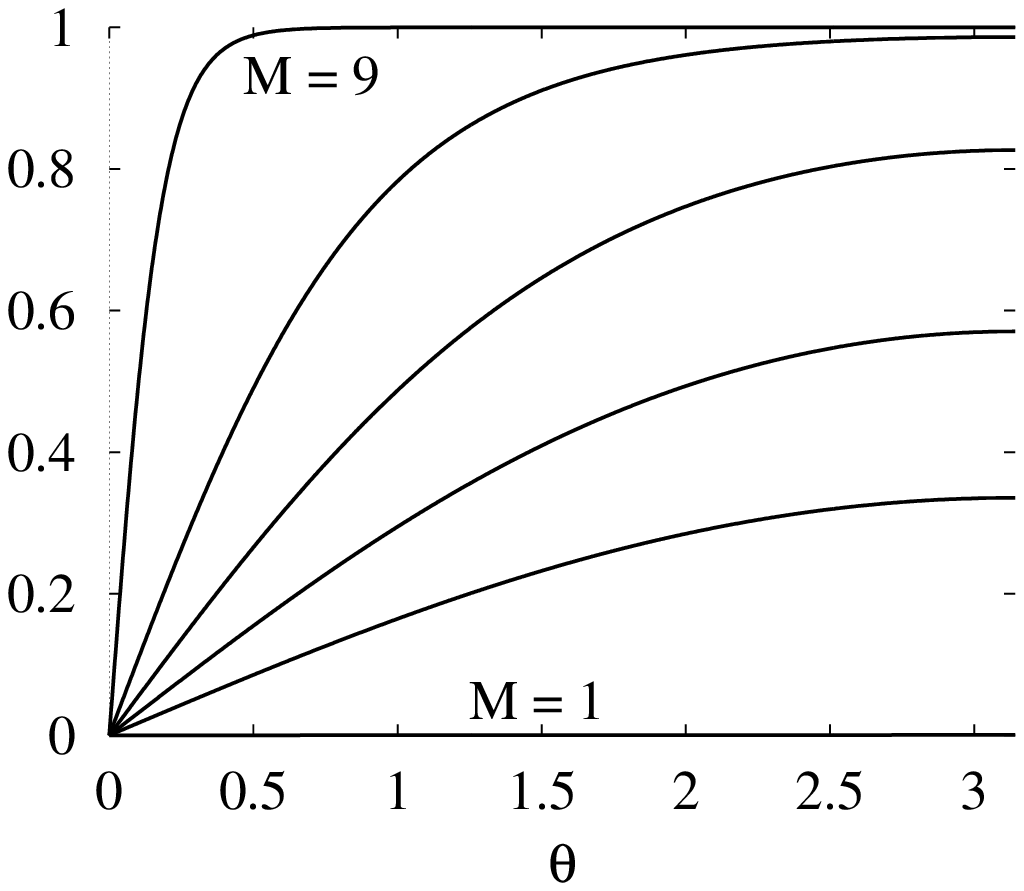,width=6.8cm} \qquad
              \epsfig{file=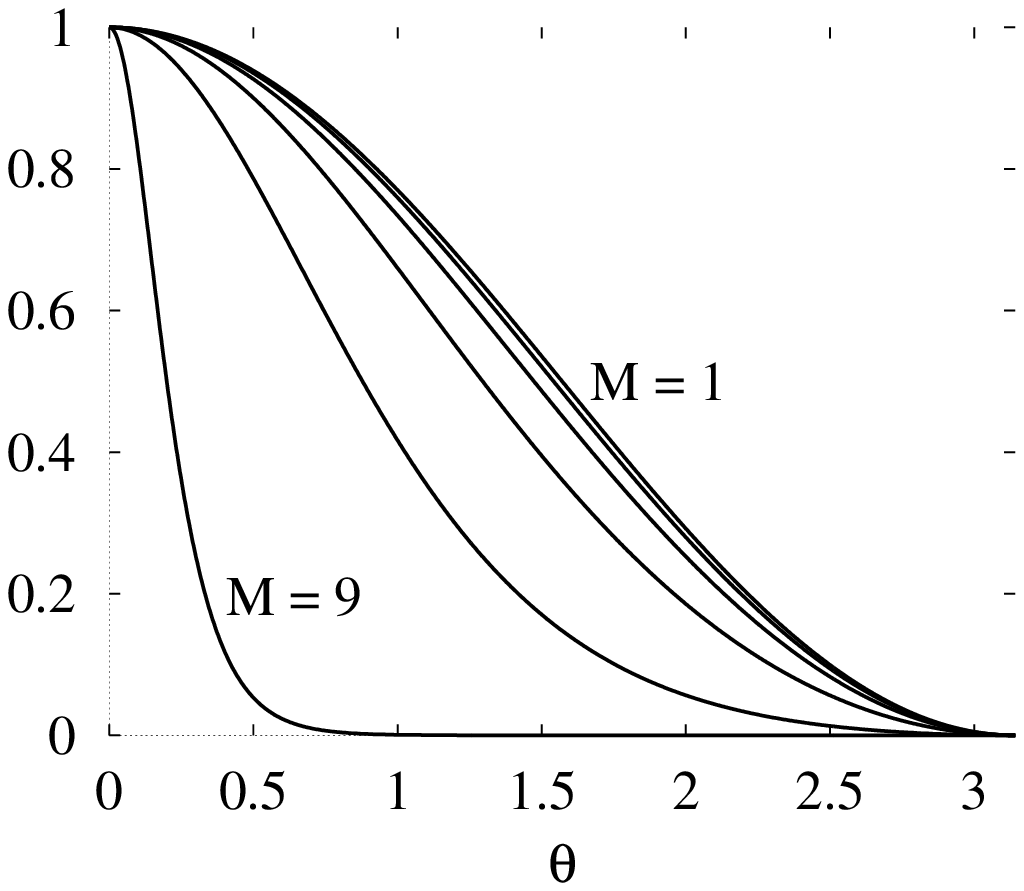,width=6.8cm}}
  \caption{Fields $X(\theta)$ (left) and $P(\theta)$ on the horizon for the
           case of a single string ending on the black hole. The string has a
           Higgs width of 1, and we plot the profiles for $\beta = N = 1$ and
           the following values of $M$: 9, 2, 1.3, 1.1, 1.03, 1.}
  \label{fig:oneprogression}
\end{figure}

\begin{figure}[htbp]
  \centerline{\epsfig{file=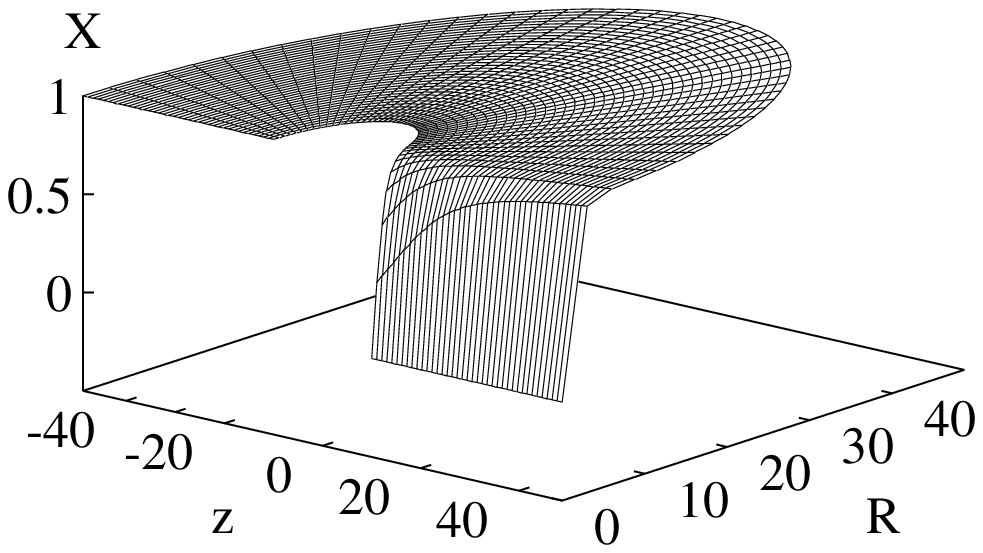,width=6.8cm} \qquad
              \epsfig{file=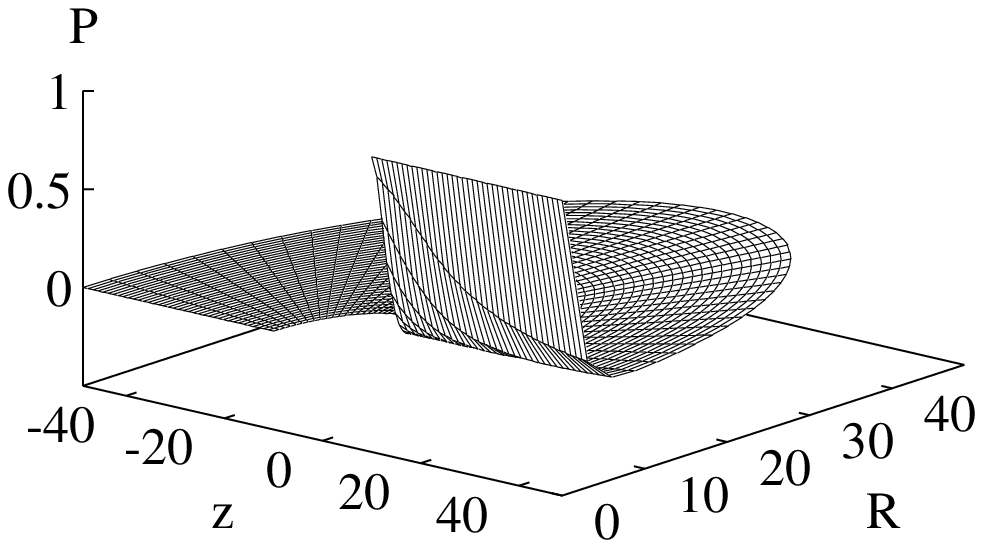,width=6.8cm}}
  \caption{Solution $X(r, \theta)$ and $P(r, \theta)$ for a single string
           ending on the black hole, and $M = Q = 10$, $\beta = N = 1$,
           $\varepsilon = 10^{-4}$, $N_r = N_\theta = 50$.}
  \label{fig:onethin}
\end{figure}

\begin{figure}[htbp]
  \centerline{\epsfig{file=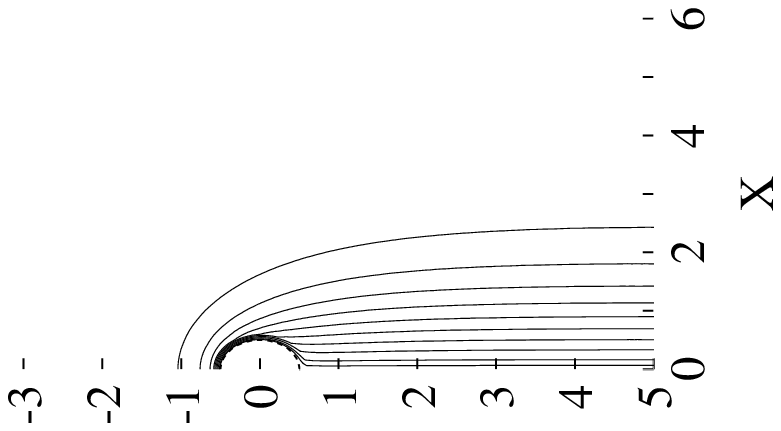,width=6.8cm,angle=270} \qquad
              \epsfig{file=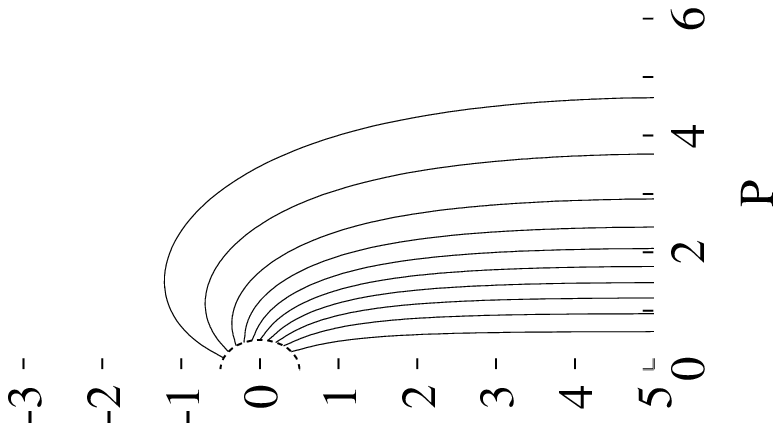,width=6.8cm,angle=270}}
  \caption{Contours of $X$ and $P$ for a single string ending on the black
           hole, and $M = Q = 1/2$, $\beta = N = 1$, $N_r = N_\theta = 100,$
           $r_m = 10$.}
  \label{fig:onesmall}
\end{figure}

On the horizon, the single string case differs from the one we have consider
previously by the boundary conditions only. At $\theta = 0$, we must clearly
have a string, but at $\theta = \pi$ nothing forces the fields to assume a
vacuum configuration. In fact, we have found that the only smooth solutions
were such that $X$ had a vanishing $\theta$-derivative at the South pole. As
figure~\ref{fig:oneprogression} shows, the value of $X_m$ then depends on the
black hole's mass, which means that if we wish to integrate the equations on
the whole grid, we also have to update the $\theta = \pi$ boundary. To find
equations of motion on this line, we assumed that $P/ \sin\theta \to 0$ and
that $X_{,\theta} \to 0$. The resulting equations, however, were particularly
unstable against numerical errors. We were finally able to tackle this problem
by artificially coupling the horizon to the rest of the grid (assuming
continuity) for some hundred iterations, and by updating the fields at $\theta
= \pi$ assuming $X_{,\theta} = P_{,\theta} = 0$ there. Examples of a thin
string solution piercing the horizon and of a thicker string being expelled are
shown on figures~\ref{fig:onethin} and \ref{fig:onesmall}.
Figure~\ref{fig:oneprogression} shows the variation of the fields on the
horizon as we thicken the string, illustrating that for $M = 1$ the Higgs field
is already expelled from the horizon.

\section{Gravitational backreaction}\label{sec:grav}

We begin this section with a lightning review of a self-gravitating
cosmic string. For convenience, we will take $N=1$ in this section.

We can readily extend the vortex to a self-gravitating system by using
Thorne's cylindrically symmetric coordinate system~\cite{TH}
\be \label{thornform}
  ds ^2 = e ^{2 \psi }dt ^2 -e ^{2( \gamma - \psi )}(dz ^2 + dR^2) -
    -  \alpha^2 e ^{-2 \psi } d \phi ^2
\ee
where $\gamma, \ \psi, \alpha$ are all functions of $R$ only.

In these coordinates, the energy momentum tensor becomes
\bml\bea
  T^0_0 &=& {\cal E} = e ^{-2(\gamma-\psi)} X ^{\prime 2} + {e^{2\psi} X^2
    P^2 \over {\alpha} ^2} + \beta e^{2(\gamma-2\psi)}
    {P ^{\prime 2} \over \alpha^2} + {\quarter} (X^2 -1) ^2 \\
  T^R_R &=&  - {\cal P}_R = - e ^{-2(\gamma-\psi)} X ^{\prime 2} +
    {e^{2\psi} X^2P^2 \over {\alpha} ^2} - \beta e^{2(\gamma-2\psi)}
    {P ^{\prime 2} \over \alpha^2} + {\quarter} (X^2 -1) ^2 \\
  T^\phi_\phi&=& -{\cal P}_\phi = e ^{-2(\gamma-\psi) } X ^{\prime 2} -
    {e^{2\psi} X^2P^2 \over {\alpha} ^2} - \beta e^{2(\gamma-2\psi)}
    {P ^{\prime 2} \over \alpha^2} + {\quarter} (X^2 -1) ^2 \\
  T^z_z &=& - {\cal P}_z = T^0_0.
\eea \eml
To zeroth order in $\epsilon = 8\pi G\eta^2$ 
introduced in section~\ref{sec:abh}
\be
  \alpha = R \;\; , \;\;\; \psi=\gamma=0 \;\; , \;\;\; X=X_0\;\;,\;\;\; P=P_0,
\ee
and conservation of the energy-momentum tensor gives
\be
  (R{\cal P}_{_0R})' = {\cal P}_{_0\phi}.
\ee

To first order in $\epsilon$ the string metric is given by~\cite{GG}
\bml \label{firstord} \bea
  {\alpha} &=& \left [1- \epsilon \int _0 ^R
    R ({\cal E}_{_0} - {\cal P}_{_0R})d R \right ] R +
    \epsilon \int _0 ^R  R ^2 ({\cal E}_{_0} - {\cal P}_{_0 R})dR \\
  \gamma &=& 2\psi= \epsilon \int _0 ^R R {\cal P}_{_0 R} dR.
\eea\eml
Where the subscript zero indicates evaluation in the flat space limit.  Since
the string functions $X$ and $P$ rapidly fall off to their vacuum values
outside the core, the integrals in~(\ref{firstord}) rapidly converge to their
asymptotic, constant, values. Writing
\be
  A = \int _0 ^\infty R ({\cal E}_{_0} - {\cal P}_{_0R})d R , \;\;\;\;
  B= \int _0 ^\infty R^2 ({\cal E}_{_0} - {\cal P}_{_0R})d R 
  \;\;\;\;\;{\rm and} \;\;\;\;\;
  C= \int _0 ^\infty R {\cal P}_{_0 R},
\ee
then the asymptotic form of the metric is
\bea \label{asycon}
  ds^2 &=& e^{\epsilon C}[dt^2-dR^2-dz^2] - R^2 (1-{\epsilon A})^2 
    e^{-{\epsilon C}} d\phi^2 \nonumber \\
  &=& d{\hat t}^2 - d{\hat R}^2 - d{\hat z}^2 - {\hat R}^2 
    (1-\epsilon (A+C))^2d\phi^2,
\eea
where ${\hat t} = e^{\epsilon C/2}t$ etc. This is a conical metric with deficit
angle
\be
  \Delta = 2\pi\epsilon (A+C) = 2\pi \epsilon\int R {\cal E}_{_0} dR
    = 8\pi G\mu,
\ee
where $\mu$ is the energy per unit length of the string.  Notice that the
deficit angle is independent of the radial stresses, but that when the radial
stresses do not vanish there is a red/blue-shift of time between infinity and
the core of the string. The only case in which these stresses do vanish is when
${\beta}=1$.

Now we are ready to consider the gravitational effect of the string
superimposed on the black hole. For this we need to consider a general static
axially symmetric solution to the Einstein-Maxwell-abelian Higgs equations
derived from the action $S_1+S_2$, where $S_1$ is given by~(\ref{abhact}), and
$S_2$ is the Einstein-Maxwell action
\be \label{emact}
  S_2 = -{1\over 16\pi G} \int d^4 x \sqrt{-g} \left [ R + 
    F_{\mu\nu}^2 \right ],
\ee
where for clarity in what follows, we have rescaled the electromagnetic
field by $\sqrt{G}$.
We may pick coordinates so that the metric takes the form
\be
  ds ^2 = e ^{2 \psi }dt ^2
    -e ^{2( \gamma - \psi )}(dz ^2 + d\rho ^2 ) -
    {\alpha}^2 e ^{-2 \psi } d \phi ^2,
\ee
where $\psi, \gamma,{\alpha}$ are independent of $t, \phi$. [Note the
deliberate similarities with the Thorne metric~(\ref{thornform}).]  We then
apply an iterative procedure to solving the field equations {\it in the 
thin string limit}, starting with the
Reissner-Nordstr{\o}m background solution, the Nielsen-Olesen forms of $X$ and
$P$, and expanding the equations of motion in terms of $\epsilon=8\pi G\eta^2$
as before. 

The usual Reissner-Nordstr{\o}m metric~(\ref{rnmetric}) is of course written in
`spherical' coordinates, whereas, in such an iterative process we require it in
axisymmetric coordinates. For future reference the coordinate transformation is
\be
  \rho = \sqrt{r^2-2Mr+Q^2} \sin\theta, \qquad
  z = (r-M)\cos\theta
\label{transfm}
\ee
and the metric and (rescaled) electromagnetic gauge potential 
(in a suitable gauge)
are given by
\be \label{rnweyl}
  ds^2 = \textstyle{{ (R_+ + R_- )^2 - 4 \Delta^2 \over
    (R_++R_-+2M)^2} } dt^2 - \textstyle{(R_++R_-+2M)^2\over 4R_+R_-}
    (d\rho^2+dz^2) - \rho^2 \textstyle{(R_++R_-+2M)^2\over
    (R_+ + R_- )^2 - 4 \Delta^2} d\phi^2,
\ee
\be \label{rnvecw}
  A_{_0\nu} = \cases{ {2Qz\over R_++R_-} \; \partial_\nu \phi & magnetic \cr
    {2Q\over(R_++R_-+2M)} \; \partial_\nu t & electric, \cr}
\ee
where
\be
  R_\pm = (z \pm \Delta)^2 + \rho^2
\ee
and $\Delta^2 = M^2-Q^2$. 

Returning to the general system, the relevant equations equations of motion
are:
\bml \label{genem} \bea
  0 &=& \partial_\rho \left ( \alpha F_\rho^{\; \nu} \right )
    + \partial_z \left ( \alpha  F_z^{\; \nu}  \right ) \label{fem} \\
  \alpha_{,zz} + \alpha_{,\rho\rho} &=& -\sqrt{-g}
    ({\cal T}^z_z + {\cal T}^\rho_\rho ) \label{alem} \\
  (\alpha\psi_{,z})_{,z} + (\alpha \psi_{,\rho})_{,\rho} &=&
    {\half}\sqrt{-g}({\cal T}^0_0 - {\cal T}^z_z 
    -{\cal T}^\rho_\rho - {\cal T}^\phi_\phi) \label{psiem} \\
%  (\alpha_{,z}^2 + \alpha_{,\rho}^2) \gamma_{,\rho} &=&
%    \sqrt{-g} ( \alpha_{,\rho} {\cal T}^z_z - \alpha_{,z}
%    {\cal T}^\rho_z) + \alpha\alpha_{,\rho}(\psi_{,\rho}^2 - \psi_{,z}^2) \\ 
%  & +& 2\alpha\alpha_{,z} \psi_{,z} \psi_{,\rho} + \alpha_{,\rho}
%    \alpha_{,\rho\rho} + \alpha_{,z} \alpha_{,z\rho} \\
%  (\alpha_{,z}^2 + \alpha_{,\rho}^2) \gamma_{,z} &=& -
%    \sqrt{-g} ( \alpha_{,z} {\cal T}^z_z + \alpha_{,\rho}
%    {\cal T}^\rho_z) - \alpha\alpha_{,z}(\psi_{,\rho}^2 - \psi_{,z}^2) \\
%  & +& 2\alpha\alpha_{,\rho} \psi_{,z} \psi_{,\rho} - \alpha_{,z}
%    \alpha_{,\rho\rho} + \alpha_{,\rho} \alpha_{,z\rho} \\
  \gamma_{,\rho\rho} + \gamma_{,zz} &=& - \psi_{,\rho}^2
    - \psi_{,z}^2 - e^{2(\gamma-\psi)} {\cal T}^\phi_\phi,
    \label{gamem}
\eea \eml
where the energy momentum tensor is given by
\be
  {\cal T}^a_b =  E^a_b + \epsilon T^a_b.
\ee
The first term is the electromagnetic contribution to the stress-energy which
is given by
\be
  E^\mu_\nu = -F_{\nu\lambda} F^{\mu\lambda} + {\quarter} F^2 \delta^\mu_\nu
\ee
and the last term is the contribution from the string, which has the explicit
form
\bml \bea
  T ^0_0 &=& V(X) + {X^2P^2\over \alpha^2e^{-2\psi}}
    + \left [ {(P_{,\rho}^2 + P_{,z}^2) \over {\beta^{-1}} \alpha^2 e^{-2\psi}}
    + (X_{,\rho}^2 + X_{,z}^2) \right ]  e^{-2(\gamma-\psi)} \\
  T^\phi_\phi &=& V(X) - {X^2P^2\over \alpha^2e^{-2\psi}}
    + \left [ - {(P_{,\rho}^2 + P_{,z}^2) \over {\beta^{-1}} \alpha^2 e^{-2\psi
    }} + (X_{,\rho}^2 + X_{,z}^2) \right ]  e^{-2(\gamma-\psi)} \\
  T^\rho_\rho + T^z_z &=&  2V(X) + {2X^2P^2\over \alpha^2e^{-2\psi}}.
\eea \eml
Note that the electromagnetic stress energy always satisfies $E^\rho_\rho +
E^z_z=0$.

As in~\cite{AGK} we now write $\alpha = \alpha_0+\epsilon\alpha_1$ etc., and
solve the Einstein-Maxwell and the string equations iteratively. The 
main difference to~\cite{AGK} is that we now have the electromagnetic
gauge potential present, which appears at O($\epsilon^0$). This means
that to O($\epsilon$) the geometry is not only affected by the string,
but also by the backreaction of the string on the electromagnetic field.

To zeroth order, we have the background solutions~(\ref{rnweyl}),
(\ref{rnvecw}), $X=X_0(R)$ and $P=P_0(R)$ where $R=r\sin\theta=\rho
e^{-\psi_0}$ as before.  Using the coordinate transformation~(\ref{transfm}),
one finds that
\be
  R^2_{,z} + R^2_{,\rho} =  {r^2\over R_+R_-}
    \left ( 1 - {2M\over r}\sin^2 \theta + {Q^2\over r^2} \sin^2\theta \right )
    \simeq e^{2(\gamma_0-\psi_0)} 
%\left ( 1 - {2M\over r}\sin^2 \theta + {Q^2\over r^2} \sin^2\theta \right )
\ee
in the core of the string, where $\sin\theta = O(M^{-1})$. Therefore, in and
near the core of the string, the relevant combinations of the zeroth order
energy momentum tensor of the string are
\bml \bea
  T^0_{_00} &=& {\cal E} + O(M^{-2})\\
  T^\phi_{_0\phi} &=& -{\cal P}_\phi + O(M^{-2})\\
  T^\rho_{_0\rho} + T^z_{_0z} &=& - ({\cal P}_R + {\cal P}_\phi ).
    \label{alrhs}
\eea \eml

As in~\cite{AGK} we will assume (and show subsequently that it is 
consistent to do so) that the perturbed solutions take the form
\be
  \alpha_1 = \rho a(R), \qquad
  \psi_1 = \psi_1(R), \qquad
  \gamma_1 = \gamma_1(R), \qquad
  A_{_1\mu} = f(R) A_{_0\mu}.
  \label{assumed}
\ee
Computing the necessary derivative of $R$ gives from~(\ref{alem}) the following
equation for $a(R)$:
\bea
  \hspace*{-1cm} \left[ 1 - {R^2\over r^2} \left ( {2M\over r} - {Q^2\over
    r^2} \right )
    \right ] a''(R) &+& \left [ {2\over R} - {Q^2\over 2r^4} - {R\over r^2}
    \left ( {3M\over r} - {Q^2\over r^2} \right ) \right ] a'(R) \nonumber \\
  &=& {1\over R^2} \left ( R^2 a'\right )' + O(M^{-2})
    = - ( {\cal E} - {\cal P}_R ) + O(M^{-2}),
\eea
which is consistently solved, as in the Schwarzschild case, by
\be
  a(R) = -\int R [ {\cal E}_{_0} - {\cal P}_{_{0R}}]dR + {1\over R}
    \int R^2 [ {\cal E}_{_0} - {\cal P}_{_{0R}}]dR
    \sim -A + {B\over R} \; {\rm as} \; R \to \infty.
\ee

Now look at the Maxwell equation for an electric potential (the magnetic
potential can be obtained by a duality transformation). Substituting in the
assumed form for the functions~(\ref{assumed}), we obtain for $f(R)$:
\be
  {f''(R)\over R} + {f'(R) \over R^2} = {\rho^2\over r^2R^2} (a'-2
  \psi_1') = O(M^{-2}),
\ee
which implies that $f=f_0$, a constant. Turning to the $\psi$ equation, and
inputting in this form of $f$, we find
\bea
  \psi_1'' \left [ 1 + {R^2\over r^2} \left ( {Q^2\over r^2}
  -{2M\over r} \right ) \right ] &+& {\psi_1'\over R} \left [
  1 - {R^2Q^2\over r^4} \right ] + {a' R\over r^2} \left (
  {Q^2\over r^2} + {M\over r} \right ) \nonumber \\
  &=& {\half} ({\cal P}_R 
  +{\cal P}_\phi) + {Q^2\over r^4} \left [ 2f_0 - 2\psi_1 \right ]
\eea
This is solved by
\be
  \psi_1 = {\half} \int R \, {\cal P}_R \sim {C\over 2} \; {\rm as} \; R \to
    \infty, \qquad f_0 = C/2
\ee
the latter value of $f_0$ being set by consistency of the $\psi$ equation
outside the core. It is then straightforward to check that $\gamma_1=2\psi_1$.
The magnetic correction is obtained either directly, or via duality, to
be 
  \be
  f_M(R) = a(R) - 2\psi_1(R) + C/2\; .
\ee

As in~\cite{AGK}, the corrections are almost the same as for the
self-gravitating cosmic string. After transforming back to Schwarzschild
coordinates, the metric outside the string core becomes
\be \label{rncorr}
  ds^2 = e^{\epsilon C}\left [ {\textstyle{\left(1 - {2M\over r} + {Q^2\over
    r^2}\right)}} dt^2 - {dr^2 \over 1 - {2M\over r} + {Q^2\over r^2}} - r^2
    d\theta^2\right] - r^2 \sin^2\theta (1-\epsilon A)^2 e^{-\epsilon C}
    d\phi^2,
\ee
where we have neglected the $B$ term from $a(R)$ since it yields a correction
$O(G\mu)\times O(E^{-1})$. The gauge potentials are
\be \label{rnveccorr}
  A_{\nu} = \cases{ Q (1-\cos\theta) [1-\epsilon(A+C/2)]\; \partial_\nu \phi & 
  magnetic, \cr
  -{Q\over r} [1+\epsilon C/2]\; \partial_\nu t & electric. \cr}
\ee
If we now rescale the metric so that $\hat t = e^{\epsilon C/2} t$, etc, (and,
accordingly, rescale the parameters $M$ and $Q$) we find
\be \label{rncorr2}
  ds^2 = {\textstyle{\left(1 - {2\hat M\over \hat r} + {\hat Q^2\over \hat
    r^2}\right)}} d\hat t^2 - {d\hat r^2 \over 1 - {2\hat M\over \hat r} + {\hat
    Q^2\over \hat r^2}} - \hat r^2 d\theta^2 - \hat r^2 \sin^2\theta
    (1-\epsilon A)^2 e^{-2\epsilon C} d\phi^2.
\ee
Again, we find a deficit angle $\Delta =2\pi\epsilon (A+C) =8 \pi G \mu$.
Besides, the gravitational mass of the black hole, $M_g$, which is given by the
coefficient of $2{\hat r}^{-1}$ in $g_{\hat t \hat t}$, has been shifted in the
presence of the string to $M_g = \hat M = e^{\epsilon C/2}M$.

On the other hand, the internal energy of the black hole (its ADM mass,
appropriately generalized to asymptotically locally flat spaces, see e.g.,
\cite{HH}) is now
\be
  M_I = \hat M (1-\epsilon A) e^{-\epsilon C}=M(1-\epsilon A) e^{-\epsilon
    C/2}.
\ee
Not only is the mass of the black hole corrected, but the physical charge of the
black hole, defined as
\be
  Q_{ph}=\cases{ {1\over 4\pi} \int_{S^2} F & magnetic, \cr
     {1\over 4\pi} \int_{S^2} *F& electric, \cr}
\ee
becomes
\be
  Q_{ph} = Q [1-\epsilon (A +C/2)].
\ee
Notice that ${M_g / Q_{ph}} >{ M_I / Q_{ph}} = {M/Q}$.

We can now write the first-order corrected solution in terms of the physical
parameters $M_I, Q_{ph}, \mu$, as
\be \label{rncorrph}
  ds^2 = {\textstyle{\left(1 - e^{4G\mu}{2M_I\over \hat r} +
    e^{8G\mu}{Q_{ph}^2\over \hat r^2}\right)}} d\hat t^2 - {d\hat r^2 \over 1 -
    e^{4G\mu}{2M_I\over \hat r} + e^{8G\mu}{Q_{ph}^2\over \hat r^2}} - \hat r^2
    d\theta^2 - e^{-4G\mu}\hat r^2 \sin^2\theta d\phi^2 ,
\ee
and
\be \label{rnveccorrph}
  A_{\nu} = \cases{ Q_{ph} (1-\cos\theta) \; \partial_\nu \phi & 
    magnetic, \cr -e^{4G\mu}{Q_{ph}\over \hat r} \; \partial_\nu \hat t &
    electric. \cr}
\ee
The corrected inner and outer horizons exterior to the core are therefore at
\be
  \hat r_{\pm} = e^{4G\mu} (M_I \pm \sqrt{M_I^2 -Q^2_{ph}}),
\ee
since we are assuming that the string is thin with respect to the back hole,
these expressions will hold except at the poles of the horizon.
Notice that the condition that the black hole is extremal (i.e., its horizon is
degenerate), is that $M_I =Q_{ph}$.  Finally, we can use the methods
of~\cite{HH} to find the entropy using the Bekenstein-Hawking formula
\be
  S={{\cal A}_h \over 4G}=16 \pi e^{4G\mu}\hat r_+^2.
\ee

Consider now an extremal black hole-cosmic string merger. If we keep the
internal energy and the charge of the black hole fixed (i.e., impose
microcanonical boundary conditions), then we find, as in~\cite{AGK}, that the
change in the gravitational mass, $\delta M_g=4G\mu M_g$ equals the energy of
the length of string swallowed by the black hole. Since the entropy increases,
we see that merging is thermodynamically favoured. And more remarkably, we see
that, at least to first order, the black hole remains extremal after the
merger.

\section{Discussion}

Our analysis in this paper appears to settle the question of whether or not
a vortex can penetrate an extremal black hole.  We have provided analytical
proofs that vortices of size smaller than a certain fraction (of order unity)
of the radius of the black hole will definitely pierce the horizon, whereas
vortices thicker than a certain lower bound will instead wrap the black hole.
The numerical analysis confirms this, and, for $N=\beta =1$, places the
transition at $M=1.8865$, or $Gm = 1.8865 / \sqrt{\lambda}\eta$.  
For a single string ending on the
black hole the Higgs field presents a similar behavior, i.e., it vanishes on
the horizon only if the black hole is small enough.  In that regime, the
magnetic field, instead of being expelled, takes the form of a monopole field.
Thus we see that single strings are always allowed to end on black holes, which
solves one of the paradoxes that the results of \cite{CCES} seemed to pose.
Finally, we have computed the backreaction effect of a thin vortex on the
geometry.  This results in the expected conical geometry, but we have been able
to check as well that the black hole remains extremal after including the
corrections to the mass and the charge.

Given that in~\cite{CEG} the expulsion of the (unbroken) magnetic field was
related to a sort of `superconducting' behavior of the extremal black hole, one
would be tempted to interpret the penetrating solutions as exhibiting the
well-known breakdown of the superconducting state for strong enough magnetic
fields.  However, this does not seem to be the case here. In~\cite{CEG} exact
solutions (which account fully for the backreaction of the gauge field) are
presented for extremal black holes in magnetic fields, and the expulsion
persists no matter how strong the magnetic fields are taken to be.

As a matter of fact, we can argue that, far from having anything to do with the
strength of the magnetic field, it is instead the presence of a mass for the
gauge vector which spoils the expulsion from the extremal horizon. In order to
illustrate this point, consider a massive vector (Proca) field, with an
explicit mass $\mu$. On the extremal horizon the field equation becomes 
\be
  \sin\theta \; \partial_\theta \left ({\partial_\theta P \over \sin \theta}
    \right ) - {\mu^2P} = 0.  
\ee 
This equation does not admit a constant solution (apart
from the trivial $P=0$). Therefore $G_{\theta\phi} \propto \partial_\theta
P_\phi \neq 0$ and we find a (locally) non-vanishing flux of the field across
the horizon.

As we have seen, things are subtler when the mass originates from
spontaneous breaking of the gauge symmetry.  The system has both massive and
massless phases, and both expulsion and penetration can be found, depending
essentially on the relative values of the vortex and black hole radii.  But the
argument above shows that the expulsion of the field can take place only if the
symmetry is {\it exactly} restored on the horizon.  It is therefore quite
remarkable that, in certain regimes, the geometry of the extremal horizon can
locally enforce the exact restoration of the symmetry.

The transition from penetration to expulsion can be viewed as a phase
transition on the horizon of the black hole.  In particular,
Fig.~\ref{fig:maxmin} is reminiscent of the behavior of, say, the magnetization
of a ferromagnet as a function of temperature, or more generally, the order
parameter of a system undergoing a second order phase transition.  The order
parameter is in this case the value of the Higgs field on the horizon, $X$, and
instead of a function of the temperature, the phase transition takes place when
we vary the (inverse) size of the horizon $\sim M^{-1}$.  Beyond a certain
critical value, $M^{-1} \geq M_c^{-1}$, the symmetry is restored throughout the
horizon.  Notice that the transition takes place when the energy scale set by
the black hole, $M$, is similar to the Higgs energy $\sqrt{\lambda}\eta$.  In a
sense, this would be the natural expectation, since $\eta$ sets the energy
scale for symmetry restoration.  But it should be stressed that this
expectation is only realized for extremal black holes: the symmetry is never
restored on non-extremal horizons\footnote{Notice as well that it would be
incorrect to think that the restoration of the symmetry comes about as an
effect of the thermal properties of black holes.  For one thing, an extremal
black hole has zero temperature.  Moreover, the thermality is only seen when
accounting for quantum effects, whereas here we work at the classical level.}.
It might be interesting to pursue this analogy further, and study, e.g.,
critical exponents near the transition point, such as $X \sim
|M-M_c|^{\tilde\beta}$ (and see, e.g., how ${\tilde\beta}$ varies, or not, at
different points on the horizon).

Another interesting question that can be explored in more generality is the
interaction between the straight cosmic string and the extremal 
black hole.
The set up we have been considering so far places the vortex in perfect
alignment with a black hole axis.  If the black hole and the vortex are
displaced relative to each other, the symmetry of the system decreases and
complication increases greatly.  We can, however, analyze in some detail the
interactions between the extremal black hole and the cosmic string when the
string is thin enough to allow us to effectively integrate out the details of
the core structure.  We can proceed in several levels of approximation.  A very
crude approximation would be taking the black hole as a test particle in the
background of a self-gravitating cosmic string, i.e., in a flat spacetime with
a conical defect.  As is well known, since the spacetime is locally flat the
test particle does not experience any force.

We can improve on this by accounting for the gravitational field of the black
hole.  In the Newtonian approximation, we would be solving the Poisson equation
for the Newtonian potential in a conical spacetime.  The effect of the conical
defect on a massive neutral particle can be understood by viewing the particle
as subject to a force coming from the `images' produced by the conical defect.
This results in the neutral particle being attracted towards the
string~\cite{S1} (see also \cite{PD}).  In contrast, when applied to a charged
`extremal particle,' such that $m=q$, this argument would yield a vanishing
force, since the gravitational and electrostatic forces between the particle
and its images cancel out.

Actually, for extremal black holes this result holds not only in the Newtonian
approximation, but also exactly in the full Einstein-Maxwell theory\cite{LE}.  
To see this, recall first that the solution for an (electric) extremal 
black hole can be written as
\bea \label{conf}
  ds^2 &=& H^{-2} dt^2 - H^2 (dx^2 +dy^2 +dz^2),\\
  A_\mu &=&(H^{-1}-1) \partial_\mu t .\nonumber
\eea
Written in this fashion, the Einstein-Maxwell equations only require that $H$
is a harmonic function in the flat $(x,y,z)$ space, $\nabla^2_{x,y,z} H
=0$. The extremal black hole is recovered by setting
\be
  H=1 +{Q\over \sqrt{(x-x_0)^2 +(y-y_0)^2 +(z-z_0)^2}},
\ee
and the horizon is at $(x,y,z)=(x_0,y_0,z_0).$\footnote{The `Schwarzschild
coordinates' used in~(\ref{xrnmetric}) correspond to setting $r - M =[(x-x_0)^2
+(y-y_0)^2 +(z-z_0)^2]^{1/2}$.} If we want to include the cosmic string, then
we just have to solve the Laplace equation for $H$, this time in a space with a
straight conical line.  The relevant solution has been given in many places,
see e.g., in the context of cosmic strings,~\cite{S1}.  In cylindrical
coordinates $(\rho,z,\phi)$ centered on the string, with conical deficit such
that $0\leq \phi \leq 2\pi/p$, if we put the black hole at $\rho=\rho_0$,
$\phi=0$ and $z=0$, then
\be \label{harm}
  H_p(z,\rho,\phi ;\rho_0)= 1+{Q\over \pi \sqrt{2\rho \rho_0}} \int_{u_o}^\infty
    {du \over \sqrt{\cosh u-\cosh u_0}} {p \sinh pu\over \cosh pu -\cos p\phi},
\ee
where $u_0$ is defined by
\be
  \cosh u_0 ={\rho^2 +z^2 +\rho_0^2 \over 2 \rho \rho_0}.
\ee
This solution is nonsingular, away from the conical line and the singularity of
the black hole. Thus, there are no forces between the extremal black hole and
the cosmic string.

At distances much larger than $\rho_0$, the harmonic function becomes
\be
  H_p \to 1+{Q p \over \sqrt{\rho^2 +z^2}}.
\ee
Since $p\approx 1+4G\mu$, we see that we reproduce the gravitational and
electric potentials in~(\ref{rncorrph}), (\ref{rnveccorrph}), with the
parameter $Q$ equaling both the internal energy $M_I$ and the physical charge
$Q_{ph}$ we had introduced earlier. The result, however, is independent of
whether the string and the black hole are merged or not. Whenever the black
hole is in the presence of a cosmic string, the lines of force are `focused,'
resulting in an increase of the gravitational and electric potentials at long
distances.

The same conclusion regarding the absence of a force between the two objects
can be reached again from another perspective. This time, neglect the
gravitational backreaction of the string, and consider a Nambu-Goto string in
the background of the extremal black hole, with metric as in~(\ref{conf}). The
static interaction potential $V(x,y,z)$ experienced by a string can be read off
from the string action, $I_{NG} \sim \int d\tau V(x,y,z)$.  If we take a
straight string along, say, the $z$ axis, this is, $T=\tau$, $Z=\sigma$,
$X,Y={\rm constant}$, then the action is
\bea
  I_{NG} &=&-\mu \int d\tau\; d\sigma\sqrt{-\det g_{\mu\nu}\partial_\alpha
    X^\mu  \partial_\beta X^\nu}\nonumber\\
  &=& -\mu L\int d\tau\ ,
\eea
($L$ is the length of the string) i.e., the static force vanishes. With a bit
more work it is easy to see that if the string is given a velocity transverse
to its axis, then its motion is slowed down as it approaches the black hole.

Clearly, in all these arguments we have been neglecting the effect of the black
hole on the string core.  But nevertheless we seem to find that, at least if
they are well separated, a straight string and an extremal black hole will
hardly feel each other's presence.  

These results appear to imply that the binding energy between the infinitely
thin string and the extremal black hole is zero.  It appears difficult to make
a clear comparison between the energy of a finite-radius vortex before and
after the merger.  On the one hand, when the vortex penetrates the horizon, a
hole is cut out from space, where a part of the string is missing.  As we have
seen in the previous section, the energy of the missing link equals the change
in the gravitational mass of the black hole.  On the other hand, since the
geometry along the string changes in the process, one should be careful about
how to define the total energy change.  In particular, since the string has an
infinite length (and hence, infinite energy), one should specify a
regularization, and choose how to fix a large but finite length of the string
before and after the merger.

We conclude by mentioning that it should be interesting to include fermion
fields and study the supersymmetry of the extreme black hole-vortex
configuration.  As is well known, in the absence of the cosmic string, the
Reissner-Nordstr{\o}m black hole can be embedded in ($N=2$) supergravity, and
the extremality condition $M=Q$ appears then as the BPS condition for the
existence of unbroken supersymmetry generators \cite{GHu}.  
Similarly, the Nielsen-Olesen
vortex with $\beta=1$ admits a supersymmetric embedding in N=2 supergravity
in three dimensions \cite{BBS}, and the solution
preserves half of the Killing spinors of the flat vacuum.  A natural question
to ask is whether the merger configuration will be supersymmetric as well.
Although we have not analyzed this point in any detail, our analysis indicates
that when the vortices are infinitely thin, the system
exhibits some of the features characteristic of BPS configurations, including
the equality of the ADM mass and the physical charge, $M_I=Q_{ph}$
(which was actually obtained for finite, if small, width strings), as well as
the absence of forces and vanishing binding energy between the two
objects\footnote{The fact that entropy is generated in the merger is known to
happen as well in BPS composite black holes, and is typically accompanied by a
reduction of the number of supersymmetry generators that are preserved.}.
Even when the vortices have a finite width, we suspect that it is also 
possible to maintain another feature of BPS systems, namely the reduction
of the equations of motion to a first order system. Although at first 
sight this seems unlikely given the lack of symmetry, it was shown in
Anderson et.\ al.\ \cite{ABGS} that for a general worldsheet embedding,
such a reduction to a first order system does occur. In this case we
have a curved geometry, nonetheless, preliminary indications are that
a generalization of this method will work.
However, the drastic change in the behavior that takes place as the vortex
grows thicker (from penetration to expulsion) would make it very surprising
that supersymmetry be present in general.

\section*{Acknowledgments}

We would like to thank Ana Ach\'ucarro, Andrew Chamblin, Konrad Kuijken, and
Andrew Sornborger for conversations, and particularly K.K.\ and A.S.\ for
making their programs freely available for comparison.  F.B.\ is supported by
an ORS award, a Swiss National Science Foundation award and a Durham University
award; R.E.\ is supported by EPSRC; and R.G.\ is supported by the Royal
Society.

\end{document}